\documentclass[12pt,onecolumn]{IEEEtran}

\usepackage{cite}
\usepackage{amsmath, amssymb, amsfonts, amsthm}
\usepackage{graphicx}
\usepackage{subcaption}
\usepackage{xcolor}
\usepackage[T1]{fontenc}
\usepackage{url}
\usepackage{enumitem}

\allowdisplaybreaks[1]

\def\BibTeX{{\rm B\kern-.05em{\sc i\kern-.025em b}\kern-.08em
    T\kern-.1667em\lower.7ex\hbox{E}\kern-.125emX}}

\newtheorem{theorem}{Theorem}

\newtheorem{lemma}{Lemma}

\newtheorem{corollary}{Corollary}

\begin{document}

\title{On the Optimum Secrecy Outage Probability and Ergodic
Secrecy Rate over Wireless Channels \\

\author{
    \IEEEauthorblockN{
        Clément Leroy\IEEEauthorrefmark{1},
        Tarak Arbi\IEEEauthorrefmark{1}, 
        Benoit Geller\IEEEauthorrefmark{1},
        Olivier Rioul\IEEEauthorrefmark{2}
    }
    \IEEEauthorblockA{\\
        \IEEEauthorrefmark{1}ENSTA, Institut Polytechnique de Paris, France\\
        \IEEEauthorrefmark{2}Telecom Paris, Institut Polytechnique de Paris, France\\
        clement.leroy@ensta-paris.fr, tarak.arbi@ensta-paris.fr, benoit.geller@ensta-paris.fr, olivier.rioul@telecom-paris.fr
    }

\thanks{This study was funded by the AID CIEDS project SEPHYTEL.}
}
}

%%%%%%%%%%%%%%%%%%%%%%%%%%%%%%%%%%%%%%%%%%%%%%%%%%%
%%                                               %%
%%                  DEBUT TEXTE                  %%
%%                                               %%
%%%%%%%%%%%%%%%%%%%%%%%%%%%%%%%%%%%%%%%%%%%%%%%%%%%

\maketitle

%%%%%%%%%%%%%%%%%%%%%%%%%%%%%%%%%%%%%%%%%%%%%%%%%%%
%%                                               %%
%%                DEBUT ABSTRACT                 %%
%%                                               %%
%%%%%%%%%%%%%%%%%%%%%%%%%%%%%%%%%%%%%%%%%%%%%%%%%%%

\begin{abstract}
    We study the secrecy of wireless channels in the presence of an eavesdropper, where the channels are random and the transmitter only has knowledge of the channel statistics. We investigate the optimal input distribution with respect to several secrecy metrics: the Secrecy Outage Probability (\textit{SOP}), defined as the probability that the coding rate $r$ exceeds the instantaneous secrecy rate; the Ergodic Secrecy Rate (\textit{ESR}), defined as the expected secrecy rate over channel realizations; and the Ergodic Positive Secrecy Rate (\textit{EPSR}), defined as the expected value of the positive part of the secrecy rate.

    We introduce two partial orderings for random channels: the \textit{uniformly less noisy} order and the \textit{less noisy on average} order. We show that when the main channel is \textit{uniformly less noisy} than the eavesdropper channel, the optimal input distribution is a non-precoded Gaussian input for both the \textit{SOP} and the \textit{EPSR}. Furthermore, we show that the same input distribution is optimal for the \textit{ESR} when the \textit{less noisy on average} order holds. In addition, similar optimality results for the \textit{SOP} and the \textit{EPSR} are obtained for single-transmit-antenna channels without requiring any channel ordering assumptions.

    Closed-form expressions of the secrecy metrics are derived for special cases of Rayleigh fading channels. 
\end{abstract}


%%%%%%%%%%%%%%%%%%%%%%%%%%%%%%%%%%%%%%%%%%%%%%%%%%%
%%                                               %%
%%                  DEBUT INTRO                  %%
%%                                               %%
%%%%%%%%%%%%%%%%%%%%%%%%%%%%%%%%%%%%%%%%%%%%%%%%%%%
\section{Introduction}

\begin{figure}[h]
    \centering
    \includegraphics[width=0.5\linewidth]{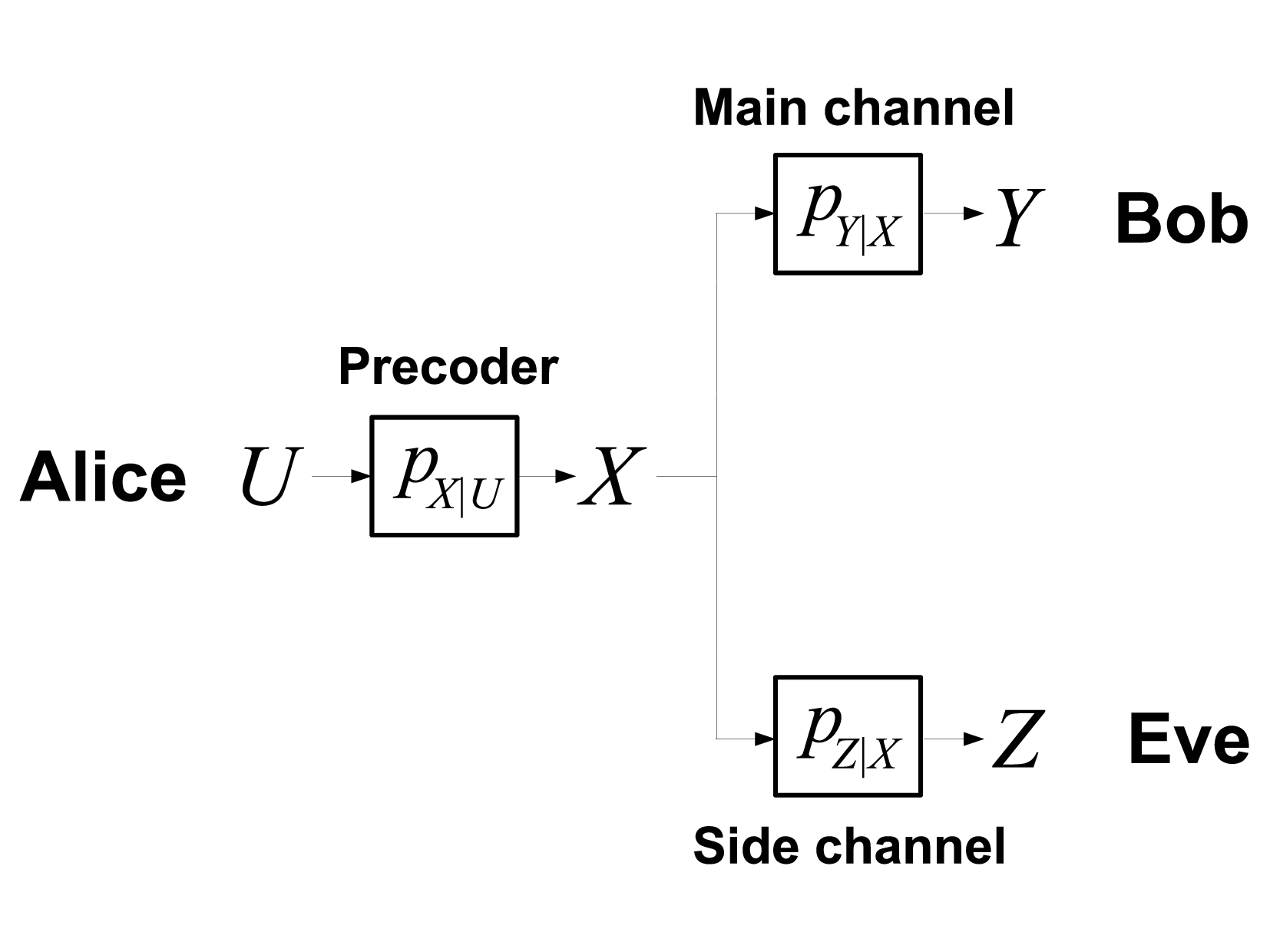}
    \caption{Illusration of the general channel model}
    \label{fig:intro}
\end{figure}

The wireless wiretap channel involves a transmitter (Alice) that emits a signal carrying a message intended to be reliably received by a legitimate receiver (Bob) while remaining confidential from an eavesdropper (Eve). Wyner's seminal work \cite{Wyner} on the degraded wiretap channel demonstrated that secrecy can be achieved through channel coding at the transmitter, with a rate up to the secrecy capacity. Csiszár and Körner \cite{Csiszar} derived the secrecy capacity for all kinds of wiretap channels, proving that it can be written as
\begin{equation}
    C_S = \max_{U - X - Y,Z} \left[ I(U;Y) - I(U;Z) \right],
    \label{eq:C_s}
\end{equation}
where \( I \) denotes the Shannon mutual information, and \( U - X - Y,Z \) represents a Markov chain in which \( U \) is an auxiliary variable, \( X \) is the transmitted signal, and \( Y \) and \( Z \) are the signals received by Bob and Eve, respectively, as illustrated in Figure \ref{fig:intro}. The maximum in \eqref{eq:C_s} is taken under an average power constraint on the signal \( X \): \( \mathbb{E}\left[ |X|^2 \right] \leq \mathcal{P}_T \), where \( \mathbb{E}[\cdot] \) denotes the expectation.

Khisti and Wornell \cite{Khisti} and Oggier and Hassibi \cite{Oggier} later independently proposed a more precise caracterization of the secrecy capacity in the MIMOME (Multiple-Input, Multiple-Output, Multiple-Eavesdropper) scenario. They showed that the secrecy capacity is achieved with a Gaussian, non-precoded input, i.e., $U \equiv X$ and \( X \) follows a Gaussian distribution, so it can be writter as a maximum over all admitible input covariance matrix:
\begin{align}
    \label{eq:C_s_mimo}
    C_S = \max_{0 \preceq \mathbf{K}_{\mathbf{x}}} \log\left( \frac{\det(\mathbf{I + \mathbf{H}\mathbf{K}_\mathbf{x}\mathbf{H}^\dagger})}{\det(\mathbf{I + \mathbf{G}\mathbf{K}_\mathbf{x}\mathbf{G}^\dagger})} \right),
\end{align}
where $\mathbf{H}$ is the legitimate channel matrix, $\mathbf{G}$ is the eavesdropper channel matrix, and $.^\dagger$ denotes the hermitian operator. The expression proposed by \cite{Khisti,Oggier} is still the result of a non-convex maximization problem. Various closed form expressions solving this optimization have later been proposed \cite{Bustin, Khina}. The I-MMSE formula \cite{GuoIMMSE} is used in \cite{Bustin} and the expression of the secrecy capacity is given as a log-det difference similar to \eqref{eq:C_s_mimo} but with matrices that have a much longer definition. The formula obtained in \cite{Khina} depends on the generalized singular values of $(\mathbf{H},\mathbf{G})$.

However, the Gaussianness of the optimal input distribution was established under the idealized and unrealistic assumption that both the legitimate channel and the eavesdropping channel are perfectly known to all parties involved in the communication. A more realistic approach is to consider that Alice only has statistical knowledge of the channels, which are modeled as random variables. This setting encompasses imperfect channel estimation, partial channel state information, and knowledge of the statistics of a fast-fading channel. Additionnaly to the secrecy capacity, the secrecy outage probability is a popular metric of secrecy in such a communication model. To the best of our knowledge, no exact expression for the secrecy capacity or the secrecy outage probability has yet been proposed in the literature for this case. Nevertheless, some bounds have been established on the secrecy capacity for channels whose randomness comes from non prefect channel estimation \cite{Rezki,Hayadi}. 

Most previous works focus on achievable secrecy rates or achievable low secrecy outage probability, typically assuming a Gaussian input that is either non-precoded~\cite{Li,Gopala,Liang,Lin} or only combined with artificial Gaussian noise~\cite{Goel,Shi,PHLin}. The assumption of the input Gaussianness is justified in~\cite{Lin} for Rayleigh MIMOME fading channels, where Bob has more antennas than Eve and the channel fading variance is larger in the main channel than in the eavesdropper channel. This assumption also rises in~\cite{Gopala, Liang} from the perfect knowledge of the instantaneous channel state by the transmitter and the receiver. 

In this paper, we investigate the optimization of three secrecy metrics involving the channel state randomness with respect to the input distribution \( p_{(U,X)} \) under an average power constraint. We show that for some large classes of wiretap channels, what we call the \textit{Ergodic Positive Secrecy Rate}, the \textit{Ergodic Serecy Rate} and the \textit{Secrecy Outage Probability} are optimized by a non-precoded Gaussian input. However, this behaviour is different for wiretap channels outside of those classes. In particular, the optimal input distribution is not necessarily Gaussian.

The remainder of this paper is organized as follows. Section \ref{sec:system_model} presents the communication model, the definitions of the secrecy metrics, and the channel classes. The optimization of the \textit{Ergodic Positive Secrecy Rate} and the \textit{Secrecy Outage Probability} is addressed in Section \ref{sec:SOP_EPSR}. The optimization of the \textit{Ergodic Secrecy Rate} is addressed in Section \ref{sec:ESR}. Section \ref{sec:rayleigh} applies the results to Rayleigh fading channels. Additional examples and counterexamples illustrating the importance of the theorem assumptions are presented in Section \ref{sec:examples}. Finally, Section \ref{sec:conclusion} concludes the paper.

\textbf{Notations:} All mutual information quantities in this paper are expressed in bits, so logarithms are taken in base~2. A complex scalar is denoted by a non-bold lowercase letter (e.g.\ $a \in \mathbb{C}$), a vector by a bold lowercase letter (e.g.\ $\mathbf{a} \in \mathbb{C}^n$), and a matrix by a bold uppercase letter (e.g.\ $\mathbf{A} \in \mathbb{C}^{n \times m}$). A random variable is denoted in the same way as one of its realizations (e.g.\ $x$ may denote a random variable in $\mathbb{C}$). The transpose of a matrix $\mathbf{A}$ is denoted by $\mathbf{A}^T$, its Hermitian by $\mathbf{A}^\dagger$, and its Frobenius norm by $\lVert \mathbf{A} \rVert$. The probability of an event in an underlying sample space $\Omega$ is denoted by $\mathbb{P}$. Expectation with respect to a random variable $x$ is denoted by $\mathbb{E}_x[\cdot]$. Variance and covariance are denoted by $\mathbb{V}[\cdot]$ and $\mathrm{cov}[\cdot]$, respectively. The minimum mean-square error (MMSE) of estimating a random variable $x$ from an observation $y$ is defined as~\cite{GuoMMSE}: $\mathbb{E}_y \big[\, \mathbb{V}_x[x | y] \,\big]$, and is denoted by $\textit{MMSE}(x | y)$. The Kullback-Leibler divergence between two distributions $p_x$ and $p_y$ is denoted by $D(x \,\|\, y)$.

%%%%%%%%%%%%%%%%%%%%%%%%%%%%%%%%%%%%%%%%%%%%%%%%%%%
%%                                               %%
%%               NOUVELLE SECTION                %%
%%                                               %%
%%%%%%%%%%%%%%%%%%%%%%%%%%%%%%%%%%%%%%%%%%%%%%%%%%%
\section{System model}
\label{sec:system_model}

\subsection{Communication model}
%\label{sec:system_model}
We consider a MIMOME (Multiple-Input, Multiple-Output, Multiple-Eavesdropper) wireless channel: the transmitter (Alice), the legitimate receiver (Bob) and the eavesdropper (Eve) have repectively $N_A$, $N_B$ and $N_E$ antennas. Alice transmits a message $u$ belonging to an arbitrary set $\mathcal{U}$, encoded into a signal $\mathbf{x}$, which is modeled as a complex random vector in $\mathbb{C}^{N_A}$. Bob observes the vector \(\mathbf{y} \in \mathbb{C}^{N_B}\), given by:
\begin{equation}
    \mathbf{y} = \mathbf{H}\mathbf{x}+\mathbf{w}_B,
    \label{eq:SIMOME_Bob}
\end{equation}
while Eve observes the vector $\mathbf{z}$ in $\mathbb{C}^{N_E}$:
\begin{equation}
    \mathbf{z} = \mathbf{G}\mathbf{x}+\mathbf{w}_E, 
    \label{eq:SIMOME_Eve}   
\end{equation}
where $\mathbf{w}_B$ and $\mathbf{w}_E$ are independent and identically distributed (i.i.d.) circularly symmetric complex Gaussian noise vectors with covariance matrix \(\mathbf{I}\). The matrix \(\mathbf{H} \in \mathbb{C}^{N_B \times N_A}\) (respectively, \(\mathbf{G} \in \mathbb{C}^{N_E} \times N_A\)) is a random matrix with distribution \(p_{\mathbf{H}}\) (respectively, \(p_{\mathbf{G}}\)). The distributions \(p_{\mathbf{H}}\) and \(p_{\mathbf{G}}\) represent the channel statistics. As the instantaneous channel states are unknown to the transmitter, the random variables $\mathbf{H}$ and $\mathbf{G}$ are independant to $u$ and $\mathbf{x}$.

\subsection{Secrecy metrics}

The secrecy of the communication system can be quantified using different metrics, each being relevant in various scenarios:
\begin{itemize}
    \item[$1)$] the \textit{Ergodic Secrecy Rate} (\textit{ESR}), defined as
    \begin{equation}
        \textit{ESR} \triangleq \mathbb{E}_{\mathbf{H},\mathbf{G}}\left[ I(u;\mathbf{y}) - I(u;\mathbf{z}) \right];
    \end{equation}

    \item[$2)$] the \textit{Ergodic Positive Secrecy Rate} (\textit{EPSR}), defined as
    \begin{equation}
        \textit{EPSR} \triangleq \mathbb{E}_{\mathbf{H},\mathbf{G}}\left[ \left( I(u;\mathbf{y}) - I(u;\mathbf{z}) \right)^+ \right];
    \end{equation}
    where $(x)^+ = \max(0,x)$;

    \item[$3)$] the \textit{Secrecy Outage Probability} (\textit{SOP}), defined for all non-negative real numbers $r$ as
    \begin{equation}
        \textit{SOP}(r) \triangleq \mathbb{P}\left[ I(u;\mathbf{y}) - I(u;\mathbf{z}) < r \right].
    \end{equation}
\end{itemize}
The \textit{ESR} represents an achievable secrecy rate when both Bob and Eve have perfect channel state information, while Alice remains unaware of it. Thus, the maximum \textit{ESR} coincides with the secrecy capacity in this specific scenario. This can be shown using the Csiszár and Körner secrecy capacity expression by considering that the channel output includes both the received signal and the instantaneous channel state:
\begin{align}
    C_S &= \max_{p_{u,\mathbf{x}}} \left[ I(u;\mathbf{y},\mathbf{H}) - I(u;\mathbf{z},\mathbf{G}) \right] \\
    &= \max_{p_{u,\mathbf{x}}} \left[ I(u;\mathbf{H}) + I(u;\mathbf{y}|\mathbf{H}) - I(u;\mathbf{G}) - I(u;\mathbf{z}|\mathbf{G}) \right] \\
    &= \max_{p_{u,\mathbf{x}}} \left[ I(u;\mathbf{y}|\mathbf{H}) - I(u;\mathbf{z}|\mathbf{G}) \right] \\
    &= \max_{p_{u,\mathbf{x}}} \left[ \mathbb{E}_{\mathbf{H},\mathbf{G}} \left[ I(u;\mathbf{y}) - I(u;\mathbf{z}) \right] \right].
\end{align}

The \textit{EPSR} corresponds to an achievable secrecy rate over slow-fading channels with feedback, where the transmitter employs an on-off power control strategy, transmitting only when the instantaneous secrecy rate is positive. This strategy is used in \cite{Gopala} to show that the secrecy capacity expression established for the slow-fading wiretap channel with feedback is indeed achievable.

The \textit{SOP} reflects the probability of information leakage to the eavesdropper when transmission occurs at a fixed secrecy rate \( r \). This metric is commonly used when the eavesdropper channel is unknown to the legitimate transmitter. It quantifies the probability that secure communication fails, i.e., that the instantaneous secrecy rate falls below the target rate \( r \).

\subsection{Partial ordering of channels}

The legitimate channel is fully characterized by the distribution \( p_{\mathbf{H}} \) of the channel matrix \( \mathbf{H} \), while the eavesdropper channel is fully characterized by the distribution \( p_{\mathbf{G}} \) of \( \mathbf{G} \). In \cite{Korner}, Körner and Marton introduced two partial orderings for noisy memoryless channels with the same input alphabet: the \textit{less noisy} order and the \textit{more capable} order. Bergmans later introduced in \cite{Bergmans} the notion of \textit{degraded} channels. These three orderings have since become classical tools for comparing communication channels. In this work, we extend these definitions to the setting of random fading channels and we propose two ways to partially order ergodic wireless channels.

\begin{itemize}
    \item Less noisy on average: in \cite{Korner}, a channel $X \rightarrow Y$ is said to be less noisy than $X \rightarrow Z$ when $I(U;Y) \geq I(U;Z)$ for every channel $U \rightarrow X$. We define the \textit{less noisy on average} order as the \textit{less noisy} order on the channels $\mathbf{x} \rightarrow(\mathbf{y},\mathbf{H})$ and $\mathbf{x} \rightarrow(\mathbf{z},\mathbf{G})$. Therefore, an ergodic channel given by $\mathbf{H}$ is said to be \textit{less noisy on average} than an ergodic channel given by $\mathbf{G}$ when:
    \begin{align}
        \forall u \rightarrow \mathbf{x}, I(u;(\mathbf{H},\mathbf{y})) \geq I(u;(\mathbf{G},\mathbf{z})),
    \end{align}
    which can also be written as:
    \begin{align}
        \label{eq:less_noisy}
        \forall u \rightarrow \mathbf{x}, \mathbb{E}_{\mathbf{H}} \left[I(u;\mathbf{y})\right] \geq \mathbb{E}_{\mathbf{G}}\left[I(u;\mathbf{z}) \right].
    \end{align}
    From \eqref{eq:less_noisy}, we see that the inequality defining the \textit{less noisy on average} order is the average over all instantaneous channel states of the inequality defining the \textit{less noisy} order. 

    \item Uniformly less noisy: we define the \textit{uniformly less noisy} order as the \textit{less noisy} order on the channels $\mathbf{x} \rightarrow \mathbf{y}$ and $\mathbf{x} \rightarrow \mathbf{z}$ for every possible instantaneous channels $(\mathbf{H}_0,\mathbf{G}_0)$. Therefore, an ergodic channel given by $\mathbf{H}$ is \textit{uniformly less noisy} than an an ergodic channel given by $\mathbf{G}$ when:
    \begin{align}
        \forall (\mathbf{H}_0, \mathbf{G}_0), \forall u \rightarrow \mathbf{x}, I(u;\mathbf{y}|\mathbf{H}=\mathbf{H}_0) \geq I(u;\mathbf{z}|\mathbf{G}=\mathbf{G}_0).
    \end{align}
    
    \item Degraded: an ergodic channel given by $\mathbf{G}$ is a degraded version of an ergodic channel given by $\mathbf{H}$ when $u - (\mathbf{x}, \mathbf{H}) - (\mathbf{y}, \mathbf{G})$ forms a Markov chain.  
\end{itemize}

It is straighforward to prove that the \textit{degraded} order is stronger than the \textit{uniformly less noisy} order, which is itself stronger than the \textit{less noisy on average} order. In other words, if an ergodic channel \( \mathbf{G} \) is a degraded version of an ergodic channel \( \mathbf{H} \), then \( \mathbf{H} \) is both uniformly less noisy and less noisy on average than \( \mathbf{G} \). Moreover, if \( \mathbf{H} \) is uniformly less noisy than \( \mathbf{G} \), then it is also less noisy on average. In this paper, we focus on the \textit{uniformly less noisy} order and the \textit{less noisy on average} order. The \textit{degraded} order is mentioned primarily because it is often easier to verify in practice whether the eavesdropper channel is a degraded version of the legitimate channel than to check whether either of the other two conditions holds.

%%%%%%%%%%%%%%%%%%%%%%%%%%%%%%%%%%%%%%%%%%%%%%%%%%%
%%                                               %%
%%               NOUVELLE SECTION                %%
%%                                               %%
%%%%%%%%%%%%%%%%%%%%%%%%%%%%%%%%%%%%%%%%%%%%%%%%%%%
\section{Secrecy Outage Probability (\textit{SOP}) and Ergodic Positive Secrecy Rate (\textit{EPSR}) optimization}
\label{sec:SOP_EPSR}

\subsection{MIMOME channel}
In this subsection, we study the general MIMOME channel, where Alice, Bob, and Eve are each equipped with an arbitrary number of antennas. We show that when the legitimate channel is \textit{uniformly less noisy} than the eavesdropper channel, both the minimum \textit{SOP} and the maximum \textit{EPSR} are achieved by a non-precoded Gaussian input. We first establish this result for the \textit{SOP}, and then use it to derive the optimality result for the \textit{EPSR}.

\begin{theorem}
    \label{th:sop_mimome}
    Let $r$ be a non-negative real number. If the ergodic channel $\mathbf{H}$ is \textit{uniformly less noisy} than the ergodic channel $\mathbf{G}$, the input distribution $p_{(u,\mathbf{x})}$ minimizing the \textit{SOP} with a fixed input covariance matrix $\mathbf{K}_\mathbf{x}$ is the one for which \(u \equiv \mathbf{x}\) and \(\mathbf{x}\) is a complex Gaussian random variable:
    \begin{equation}
        \label{eq:sop_mimome}
        \min_{\substack{p_{(u,\mathbf{x})} \\ \mathrm{cov}[\mathbf{x}]=\mathbf{K}_\mathbf{x}}} \textit{SOP}(r) = \mathbb{P} \left[\log\left( \frac{\det (\mathbf{I} + \mathbf{H}^\dagger \mathbf{K}_{\mathbf{x}} \mathbf{H})}{\det (\mathbf{I} + \mathbf{G}^\dagger \mathbf{K}_{\mathbf{x}} \mathbf{G})} \right) < r\right].
    \end{equation}
\end{theorem}

Note that in \ref{th:sop_mimome}, the matrix $\mathbf{K}_\mathbf{x}$ also fixes the total input power as $\mathrm{Tr}(\mathbf{K}_\mathbf{x})=\mathcal{P}_T$. In order to prove theorem \ref{th:sop_mimome}, we first establish the following lemma. 

\begin{lemma}
    \label{lem:more_noisy_maximization}
    Given that the instantaneous wireless channel $\mathbf{H}_0$ is less noisy than the instantaneous channel $\mathbf{G}_0$, and that $\mathrm{cov}(\mathbf{x})$ is fixed to $\mathbf{K}_{\mathbf{x}}$, we have:
    \begin{align}
        \max_{\substack{p_{(u,\mathbf{x})} \\ \mathrm{cov}[\mathbf{x}]=\mathbf{K}_\mathbf{x}}} \left[ I(u;\mathbf{y}) - I(u;\mathbf{z})\right] &= I(\mathbf{x}_{\text{g}};\mathbf{y}) - I(\mathbf{x}_{\text{g}};\mathbf{z})
    \end{align} 
    where $\mathbf{x}_{\text{g}}$ is the Gaussian random variable of covariance matrix $\mathbf{K}_\mathbf{x}$.
\end{lemma}

\begin{proof}[Proof of Lemma \ref{lem:more_noisy_maximization}]
    Since the instantaneous channel $\mathbf{H}_0$ is less noisy than the instantaneous channel $\mathbf{G}_0$, we know by \cite{Csiszar}: 
    \begin{align}
        \label{eq:lem_sop_mimome_1}
        \max_{\substack{p_{(u,\mathbf{x})} \\ \mathrm{cov}[\mathbf{x}]=\mathbf{K}_\mathbf{x}}} \left[ I(u;\mathbf{y}) - I(u;\mathbf{z})\right] = \max_{\substack{p_{(u,\mathbf{x})} \\ \mathrm{cov}[\mathbf{x}]=\mathbf{K}_\mathbf{x}}} \left[ I(\mathbf{x};\mathbf{y}) - I(\mathbf{x};\mathbf{z})\right].
    \end{align}
    Moreover, using the classical \textit{less noisy} order on instantaneous channels again, we know by \cite{VanDijk} that the function 
    \begin{align}
        p_\mathbf{x} \mapsto I(\mathbf{x};\mathbf{y}) - I(\mathbf{x};\mathbf{z})
    \end{align}
    is concave in the input distribution $p_\mathbf{x}$. Then, using the decomposition proposed by Pinsker \cite{PinskerRusse,Pinsker} we get:
    \begin{align}
        &I(\mathbf{x};\mathbf{y}) - I(\mathbf{x};\mathbf{z}) \nonumber\\
        &= \left( I(\mathbf{x}_{\text{g}};\mathbf{y}_{\text{g}}) + D(\mathbf{w}\,\|\,\mathbf{w}_{\text{g}}) - D(\mathbf{y}\,\|\,\mathbf{y}_{\text{g}}) \right) - \left( I(\mathbf{x}_{\text{g}};\mathbf{z}_{\text{g}}) + D(\mathbf{w}\,\|\,\mathbf{w}_{\text{g}}) - D(\mathbf{z}\,\|\,\mathbf{z}_{\text{g}}) \right) \\
        &= (I(\mathbf{x}_{\text{g}};\mathbf{y}_{\text{g}}) - I(\mathbf{x}_{\text{g}};\mathbf{z}_{\text{g}}))+ (D(\mathbf{z}\,\|\,\mathbf{z}_{\text{g}}) - D(\mathbf{y}\,\|\,\mathbf{y}_{\text{g}})), \label{eq:lem_sop_mimome_2_step2}
    \end{align}
    where the subscript "$\text{g}$" indicates that the variable is Gaussian with the same covariance as the original random variable. When the covariance matrix \(\mathbf{K}_{\mathbf{x}}\) of \(\mathbf{x}\) is fixed, the first term in \eqref{eq:lem_sop_mimome_2_step2} is constant, so the second term is concave in the input distribution \(p_{\mathbf{x}}\). Moreover, both \(D(\mathbf{z}\,\|\,\mathbf{z}_{\text{g}})\) and \(D(\mathbf{y}\,\|\,\mathbf{y}_{\text{g}})\) are 0 when \(\mathbf{x} = \mathbf{x}_{\text{g}}\), which is their minimum. Therefore, \(p_{\mathbf{x}_{\text{g}}}\) is a stationnary point of the second term in \eqref{eq:lem_sop_mimome_2_step2}. Since it is also concave, \eqref{eq:lem_sop_mimome_2_step2} achieves its maximum with \(p_{\mathbf{x}_{\text{g}}}\). Therefore, the maximum in \eqref{eq:lem_sop_mimome_1} is also achieved with \(p_{\mathbf{x}_{\text{g}}}\).
\end{proof}

\begin{proof}[Proof of Theorem \ref{th:sop_mimome}]
    Minimizing the \textit{SOP} is equivalent to maximizing 
    \begin{align}
        1-\textit{SOP}(r) = \mathbb{P} \left[ I(u;\mathbf{y}) - I(u;\mathbf{z}) \geq r \right].
    \end{align}
    It is assumed that the channel given by $\mathbf{H}$ is uniformly less noisy than the channel given by $\mathbf{G}$ so Lemma \ref{lem:more_noisy_maximization} can be used for every instantaneous channel $(\mathbf{H}_0,\mathbf{G}_0)$. We get that the maximizing input of $I(u;\mathbf{y}) - I(u;\mathbf{z})$ is the non precoded Gaussian input and is the same for every possible instantaneous channels. Therefore, when the covariance matrix of $\mathbf{x}$ is fixed, the \textit{SOP} is minimized by the non-precoded Gaussian input. Finally, to obtain \eqref{eq:sop_mimome}, we have to notice that
    \begin{align}
        I(\mathbf{x}_{\text{g}};\mathbf{y}_{\text{g}}) - I(\mathbf{x}_{\text{g}};\mathbf{z}_{\text{g}}) = \log\left( \frac{\det (\mathbf{I} + \mathbf{H}^\dagger \mathbf{K}_{\mathbf{x}} \mathbf{H})}{\det (\mathbf{I} + \mathbf{G}^\dagger \mathbf{K}_{\mathbf{x}} \mathbf{G})} \right).
    \end{align}
\end{proof}

The optimization of the \textit{EPSR} given in the following theorem directly follows from the minimization of the \textit{SOP} in Theorem \ref{th:sop_mimome}.

\begin{theorem}
    \label{th:epsr_mimome}
    If the ergodic channel $\mathbf{H}$ is \textit{uniformly less noisy} than the ergodic channel $\mathbf{G}$, the input distribution \(p_{(u,\mathbf{x})}\) with a fixed input covariance matrix $\mathbf{K}_\mathbf{x}$ that maximizes the Ergodic Positive Secrecy Rate is the one for which \(u \equiv \mathbf{x}\) and \(\mathbf{x}\) is a complex Gaussian random variable:
    \begin{equation}
        \label{eq:epsr_mimome}
        \max_{\substack{p_{(u,\mathbf{x})} \\ \mathrm{cov}[\mathbf{x}]=\mathbf{K}_\mathbf{x}}} \textit{EPSR} = \mathbb{E}_{h,g} \left[ \log\left( \frac{\det (\mathbf{I} + \mathbf{H}^\dagger \mathbf{K}_{\mathbf{x}} \mathbf{H})}{\det (\mathbf{I} + \mathbf{G}^\dagger \mathbf{K}_{\mathbf{x}} \mathbf{G})} \right)\right].
    \end{equation}
\end{theorem}

\begin{proof}[Proof of Theorem \ref{th:epsr_mimome}]
    \begin{align}
        \max_{\substack{p_{(u,\mathbf{x})} \\ \mathrm{cov}[\mathbf{x}]=\mathbf{K}_\mathbf{x}}} \mathbb{E}_{\mathbf{H},\mathbf{G}}\left[\left(I(u;\mathbf{y}) - I(u;\mathbf{z})\right)^+\right] 
        &= \max_{\substack{p_{(u,\mathbf{x})} \\ \mathrm{cov}[\mathbf{x}]=\mathbf{K}_\mathbf{x}}} \int_0^{+\infty} \mathbb{P}\left[\left(I(u;\mathbf{y}) - I(u;\mathbf{z})\right)^+ \geq r\right]\, dr \label{eq:epsr_mimome_proof_step1}\\
        & \leq \int_0^{+\infty} \max_{p_{(u,\mathbf{x})}} \mathbb{P}\left[\left(I(u;\mathbf{y}) - I(u;\mathbf{z})\right)^+ \geq r\right] \, dr \label{eq:epsr_mimome_proof_step3}\\
        &= \int_0^{+\infty}  \mathbb{P}\left[I(\mathbf{x}_{\text{g}};\mathbf{y}) - I(\mathbf{x}_{\text{g}};\mathbf{z}) \geq r\right] \, dr \label{eq:epsr_mimome_proof_step4}\\
        &= \mathbb{E}_{\mathbf{H},\mathbf{G}}\left[\left(I(\mathbf{x}_{\text{g}};\mathbf{y}) - I(\mathbf{x}_{\text{g}};\mathbf{z})\right)\right] \label{eq:epsr_mimome_proof_step5}\\
        &= \mathbb{E}_{\mathbf{H},\mathbf{G}} \left[ \log\left( \frac{\det (\mathbf{I} + \mathbf{H}^\dagger \mathbf{K}_{\mathbf{x}} \mathbf{H})}{\det (\mathbf{I} + \mathbf{G}^\dagger \mathbf{K}_{\mathbf{x}} \mathbf{G})} \right) \right],
    \end{align}
    where \eqref{eq:epsr_mimome_proof_step1} and \eqref{eq:epsr_mimome_proof_step5} follow from the classical formula of the expectation of non negative random variables, and \eqref{eq:epsr_mimome_proof_step4} follows from Theorem \ref{th:sop_mimome} and the fact that the difference of the mutual informations is always non-negative due to the \textit{uniformly less noisy} order between both channels. 
\end{proof}

In section \ref{subsec:degenerated}, we show that the optimality of the non-precoded Gaussian input obtained in Theorems \ref{th:sop_mimome} and \ref{th:epsr_mimome} does not hold without the \textit{uniformly less noisy} order assumption.

\subsection{SIMOME channel}
In this subsection, we prove that the optimal input distribution for the \textit{SOP} and the \textit{EPSR} is still a non-precoded Gaussian distribution when $N_A=1$ and both receivers have an arbitrary number of antenna. This scenario is refered to as a SIMOME (Single Input, Multiple Output, Multipe Eavesdropper) channel. Contrary to the MIMOME case, no order between channels is necessary to obtain the optimality of the non-precoded Gaussian input.

\begin{theorem}
    \label{th:sop_simome}
    Let $r$ be a non-negative real number. If $N_A=1$ (SIMOME scenario), the input distribution $p_{(u,x)}$ minimizing the Secrecy Outage Probability under the average power constraint $\mathcal{P}_T$ is the one for which \(u \equiv x\) and \(x\) is a complex Gaussian random variable with variance \(\mathcal{P}_T\); that is:
    \begin{equation}
        \min_{\substack{p_{(u,x)} \\ \mathbb{E}[|x|^2] \leq \mathcal{P}_T}} \textit{SOP}(r) = \mathbb{P} \left[\log\left( \frac{1+\mathcal{P}_T \lVert \mathbf{h} \rVert ^2}{1+\mathcal{P}_T \lVert \mathbf{g} \rVert ^2} \right) < r\right],
    \end{equation}
    where $\mathbf{h}$ (resp. $\mathbf{g}$) is the legitimate (resp. eavesdropper) channel matrix of dimension $1 \times N_B$ (resp. $1 \times N_E$).
\end{theorem}

\begin{proof}[Proof of Theorem \ref{th:sop_simome}]
    Minimizing the \textit{SOP} is equivalent to maximizing 
    \begin{align}
        1-\textit{SOP}(r) = \mathbb{P} \left[ I(u;\mathbf{y}) - I(u;\mathbf{z}) \geq r \right].
    \end{align}
    We now simplify the SIMOME expression into a SISOSE (Single Input, Single Output, Single Eavesdropper) equivalent expression using basic matrix transformations. Let us first define $\mathbf{\Phi}_\mathbf{h}$ and $\mathbf{\Phi}_\mathbf{g}$ as unitary matrices such that the first column of $\mathbf{\Phi}_\mathbf{h}$ is $\frac{\mathbf{h}}{\lVert \mathbf{h} \rVert}$ and the first column of $\mathbf{\Phi}_\mathbf{g}$ is $\frac{\mathbf{g}}{\lVert \mathbf{g} \rVert}$. We have: $\mathbf{h} = \mathbf{\Phi}_\mathbf{h} .(\lVert \mathbf{h} \rVert \space{} 0 ...0)^T$ and $\mathbf{g} = \mathbf{\Phi}_\mathbf{g} .(\lVert \mathbf{g} \rVert \space{} 0 ...0)^T$. Thus, we have:
    \begin{align}
        I(u;\mathbf{y}) - I(u;\mathbf{z})
        &= I(u;\mathbf{h}x+\mathbf{w}_B) -I(u;\mathbf{g}x+\mathbf{w}_E) \\
        &=I(u;\mathbf{\Phi}_\mathbf{h} .(\lVert \mathbf{h} \rVert \space{} 0 ...0)^Tx+\mathbf{w}_B) -I(u;\mathbf{\Phi}_\mathbf{g} .(\lVert \mathbf{g} \rVert \space{} 0 ...0)^Tx+\mathbf{w}_E) \\
        &=I(u;(\lVert \mathbf{h} \rVert \space{} 0 ...0)^Tx+\mathbf{\Phi}_\mathbf{h}^T.\mathbf{w}_B) -I(u;(\lVert \mathbf{g} \rVert \space{} 0 ...0)^Tx+\mathbf{\Phi}_\mathbf{g}^T.\mathbf{w}_E) \\
        &=I(u;\lVert \mathbf{h} \rVert x+w_B) -I(u;\lVert \mathbf{g} \rVert x+w_E),
    \end{align}
    where \(w_B\) and \(w_E\) are unit-variance circularly symmetric complex Gaussian noise.

    By \cite{SLYC}, we know that if $\lVert \mathbf{h} \rVert \geq \lVert \mathbf{g} \rVert$, then
    \begin{align}
        \max_{p_{(u,x)}} I(u;\mathbf{y}) - I(u;\mathbf{z}) = I(x_{\text{g}};\mathbf{y}) - I(x_{\text{g}};\mathbf{z}),
    \end{align}
    where $x_{\text{g}}$ is a circularly symmetric complex Gaussian random variable with variance $\mathcal{P}_T$.

    Besides, if $\lVert \mathbf{h} \rVert \leq \lVert \mathbf{g} \rVert$, then by the Data-Processing Inequality (DPI), for all $p_{(u,x)}$, we have
    \begin{align}
        I(u;\mathbf{y}) - I(u;\mathbf{z}) \leq 0.
    \end{align}

    Finally, we have
    \begin{align}
        \mathbb{P} \left[ I(u;\mathbf{y}) - I(u;\mathbf{z}) \geq r \right] \nonumber
        &= \mathbb{P} \left[ (I(u;\mathbf{y}) - I(u;\mathbf{z}) \geq r) \cap (\lVert \mathbf{h} \rVert \geq \lVert \mathbf{g} \rVert)\right] \\
        &\leq \mathbb{P} \left[ (I(x_{\text{g}};\mathbf{y}) - I(x_{\text{g}};\mathbf{z}) \geq r) \cap (\lVert \mathbf{h} \rVert \geq \lVert \mathbf{g} \rVert)\right] \\
        &= \mathbb{P} \left[ I(x_{\text{g}};\mathbf{y}) - I(x_{\text{g}};\mathbf{z}) \geq r\right].
    \end{align}
    To conclude the proof, we are left to notice that 
    \begin{align}
        I(\mathbf{x}_{\text{g}};\mathbf{y}) - I(\mathbf{x}_{\text{g}};\mathbf{z}) = \log\left( \frac{1+\mathcal{P}_T \lVert \mathbf{h} \rVert ^2}{1+\mathcal{P}_T \lVert \mathbf{g} \rVert ^2} \right).
    \end{align}
\end{proof}

The reasoning used to prove Theorem~\ref{th:epsr_mimome} can be directly extended to derive the optimal \textit{EPSR} for the SIMOME channel.

\begin{theorem}\label{th:epsr_simome}
    If $N_A=1$ (SIMOME scenario), the input distribution \(p_{(u,x)}\) that maximizes the Ergodic Positive Secrecy Rate under the average power constraint \(\mathcal{P}_T\) is the one for which \(u \equiv x\) and \(x\) is a complex Gaussian random variable with variance \(\mathcal{P}_T\); that is:
    \begin{equation}
        \max_{\substack{p_{(u,x)} \\ \mathbb{E}[|x|^2] \leq \mathcal{P}_T}} \textit{EPSR} = \mathbb{E}_{\mathbf{h},\mathbf{g}} \left[ \log\left( \frac{1+\mathcal{P}_T \lVert \mathbf{h} \rVert^2}{1+\mathcal{P}_T \lVert \mathbf{g} \rVert^2} \right)^+\right].
    \end{equation}
\end{theorem}

\begin{proof}
    The proof for Theorem \ref{th:epsr_simome} is the same as the proof for Theorem \ref{th:epsr_mimome} but using the result of Theorem \ref{th:sop_simome} instead of Theorem \ref{th:sop_mimome}.
\end{proof}

%%%%%%%%%%%%%%%%%%%%%%%%%%%%%%%%%%%%%%%%%%%%%%%%%%%
%%                                               %%
%%               NOUVELLE SECTION                %%
%%                                               %%
%%%%%%%%%%%%%%%%%%%%%%%%%%%%%%%%%%%%%%%%%%%%%%%%%%%
\section{Ergodic Secrecy Rate (\textit{ESR}) optimization}
\label{sec:ESR}

\subsection{MIMOME channel}
In this subsection, we study the general MIMOME channel: Alice, Bob and Eve all have an arbitrary number of antenna. We show that when the legitimate channel is \textit{less noisy on average} than the eavesdropper channel, the maximal \textit{ESR} is achieved by a non-precoded Gaussian input. 

\begin{theorem}
\label{th:esr_mimome}
    If the legitimate channel is less noisy on average than the eavesdropper channel, then the ESR is maximized by a non-precoded Gaussian input distribution:
    \begin{equation}
        \max_{\substack{p_{(u,\mathbf{x})} \\ \mathrm{cov}\left[\mathbf{x}\right]=\mathbf{K}_\mathbf{x}}} \textit{ESR} = \mathbb{E}_{\mathbf{H},\mathbf{G}} \left[ \log\left( \frac{\det(\mathbf{I}+ \mathbf{H}^\dagger \mathbf{K}_\mathbf{x} \mathbf{H})}{\det(\mathbf{I}+ \mathbf{G}^\dagger \mathbf{K}_\mathbf{x} \mathbf{G})} \right) \right].
    \end{equation}
\end{theorem}

\begin{proof}[Proof of Theorem \ref{th:esr_mimome}]
    We can decompose the \textit{ESR} as
    \begin{align}
        \textit{ESR} &= \mathbb{E}_{\mathbf{H},\mathbf{G}} \left[ I(u;\mathbf{y}) - I(u; \mathbf{z})\right] \\
        &= \mathbb{E}_{\mathbf{H},\mathbf{G}} \left[ I(\mathbf{x};\mathbf{y}) - I(\mathbf{x}; \mathbf{z}) \right] - \mathbb{E}_{\mathbf{H},\mathbf{G}} \left[I(\mathbf{x};\mathbf{y}|u) - I(\mathbf{x}; \mathbf{z}|u)\right]. \label{eq:esr_mimome_proof_part1_step2}
    \end{align}
    As the legitimate channel is \textit{less noisy on average} than the eavesdropper channel, the second term in \eqref{eq:esr_mimome_proof_part1_step2} is non negative and is zero when $u \equiv \mathbf{x}$. Thus, maximizing the \textit{ESR} reduces to the maximization of the first term in \eqref{eq:esr_mimome_proof_part1_step2}.

    Then, as the legitimate channel is \textit{less noisy on average} than the eavesdropper channel, the \textit{ESR} is non negative for every input. From \eqref{eq:esr_mimome_proof_part1_step2}, we obtain for every $u \rightarrow \mathbf{x}$:
    \begin{align}\label{eq:esr_mimome_proof_part2}
        \mathbb{E}_{\mathbf{H},\mathbf{G}} \left[ I(\mathbf{x};\mathbf{y}) - I(\mathbf{x}; \mathbf{z}) \right] \geq \mathbb{E}_{\mathbf{H},\mathbf{G}} \left[I(\mathbf{x};\mathbf{y}|u) - I(\mathbf{x}; \mathbf{z}|u)\right].
    \end{align}
    Inequality \eqref{eq:esr_mimome_proof_part2} is exactly the concavity caracterization of the fonction
    \begin{align}
        p_\mathbf{x} \mapsto \mathbb{E}_{\mathbf{H},\mathbf{G}} \left[ I(\mathbf{x};\mathbf{y}) - I(\mathbf{x}; \mathbf{z}) \right],
    \end{align}
    whose expression can further be decomposed using the Pinsker formula \cite{PinskerRusse,Pinsker}:
    \begin{align}
        &\mathbb{E}_{\mathbf{H},\mathbf{G}} \left[I(\mathbf{x};\mathbf{y}) - I(\mathbf{x};\mathbf{z}) \right] \nonumber\\
        &=  \mathbb{E}_{\mathbf{H},\mathbf{G}} \left[ I(\mathbf{x}_{\text{g}};\mathbf{y}_{\text{g}}) + D(\mathbf{w}\,\|\,\mathbf{w}_{\text{g}}) - D(\mathbf{y}\,\|\,\mathbf{y}_{\text{g}}) \right] -  \mathbb{E}_{\mathbf{H},\mathbf{G}} \left[ I(\mathbf{x}_{\text{g}};\mathbf{z}_{\text{g}}) + D(\mathbf{w}\,\|\,\mathbf{w}_{\text{g}}) - D(\mathbf{z}\,\|\,\mathbf{z}_{\text{g}}) \right] \\
        &=  \mathbb{E}_{\mathbf{H},\mathbf{G}} \left[I(\mathbf{x}_{\text{g}};\mathbf{y}_{\text{g}}) - I(\mathbf{x}_{\text{g}};\mathbf{z}_{\text{g}}) \right] +  \mathbb{E}_{\mathbf{H},\mathbf{G}} \left[ D(\mathbf{z}\,\|\,\mathbf{z}_{\text{g}}) - D(\mathbf{y}\,\|\,\mathbf{y}_{\text{g}}) \right], \label{eq:esr_mimome_part3_step2}
    \end{align}
    where the subscript "$\text{g}$" indicates that the variable is Gaussian with the same covariance as the original random variable. When the covariance matrix \(\mathbf{K}_{\mathbf{x}}\) of \(\mathbf{x}\) is fixed, the first term in \eqref{eq:esr_mimome_part3_step2} is constant, so the second term is concave in the input distribution \(p_{\mathbf{x}}\). Moreover, both \(\mathbb{E}_{\mathbf{H},\mathbf{G}} \left[D(\mathbf{z}\,\|\,\mathbf{z}_{\text{g}})\right]\) and \(\mathbb{E}_{\mathbf{H},\mathbf{G}} \left[D(\mathbf{y}\,\|\,\mathbf{y}_{\text{g}})\right]\) are 0 when \(\mathbf{x} = \mathbf{x}_{\text{g}}\), which is their minimum. Therefore, \(p_{\mathbf{x}_{\text{g}}}\) is a stationnary point of the second term in \eqref{eq:esr_mimome_part3_step2}. Since this term is also concave in the input distribution, the maximum of the \textit{ESR} is achieved with \(p_{\mathbf{x}_{\text{g}}}\). 

    The straightforward derivation of the mutual informations with Gaussian inputs concludes the proof.
\end{proof}

In the proof of Theorem~\ref{th:esr_mimome}, the \textit{less noisy on average} order plays a central role. This assumption is essential and cannot be relaxed, as demonstrated by the counterexample in Section~\ref{subsec:degenerated}, where the order fails and the maximum \textit{ESR} is not achieved by a non-precoded Gaussian input.

\subsection{SIMOME channel}
Unlike the optimization of the \textit{SOP} and the \textit{EPSR}, where reducing the problem to the SIMOME case leads to a unique general solution, the optimization of the \textit{ESR} is not significantly simplified in this scenario.

\begin{theorem}\label{th:esr_simome}
    Let us consider the case $N_A=1$ (SIMOME). Let \(c_{x|u}\) denote a fixed precoder distribution (the distribution of \(x\) given \(u\)), and let \(p_{(u,x)} = p_u c_{x|u}\) and \(q_{(u,x)} = q_u c_{x|u}\) be two input distributions sharing the same precoder, such that \(\mathbb{E}\!\left[ |x|^2 \right] = \mathcal{P}_T\). Denote by \textit{ESR}\((p_{(u,x)}, p_{\mathbf{h}}, p_{\mathbf{g}})\) (respectively, \textit{ESR}\((q_{(u,x)}, p_{\mathbf{h}}, p_{\mathbf{g}})\)) the \textit{ESR} obtained with the input distribution \(p_{(u,x)}\) (respectively, \(q_{(u,x)}\)) and the channel statistics \(p_{\mathbf{h}}\) and \(p_{\mathbf{g}}\). Exactly one of the following two statements holds:
    \begin{itemize}
        \item[(i)] For all channel statistics \(p_{\mathbf{h}}\) and \(p_{\mathbf{g}}\),
            \begin{equation}
                \label{eq:th_esr_trivial}
                \textit{ESR}(p_{(u,x)}, p_{\mathbf{h}}, p_{\mathbf{g}}) = \textit{ESR}(q_{(u,x)}, p_{\mathbf{h}}, p_{\mathbf{g}}).
            \end{equation}
        \item[(ii)] There exist channel statistics \(p_{\mathbf{h}}\) and \(p_{\mathbf{g}}\) such that
            \begin{equation}
                \label{eq:th_esr}
                \textit{ESR}(p_{(u,x)}, p_{\mathbf{h}}, p_{\mathbf{g}}) < \textit{ESR}(q_{(u,x)}, p_{\mathbf{h}}, p_{\mathbf{g}}).
            \end{equation}
    \end{itemize}
\end{theorem}

In Theorem~\ref{th:esr_simome}, statement \textit{(i)} corresponds to the trivial case in which the distributions \(p_{(u,x)}\) and \(q_{(u,x)}\) yield the same overall behavior. This occurs, for example, when \(p_{(u,x)} = q_{(u,x)}\).

Moreover, since the distributions \(p_{(u,x)}\) and \(q_{(u,x)}\) play symmetrical roles in statement \textit{(ii)} of Theorem~\ref{th:esr_simome}, they can be interchanged. Thus, no single distribution maximizes the \textit{ESR} in general, as the optimal distribution strongly depends on the channel statistics. In particular, the Gaussian input is not optimal in every scenario. Two illustrative examples are provided in Section~\ref{subsec:esr_example}: one where BPSK modulation outperforms Gaussian coding, and another where Gaussian coding achieves better performance.

Besides, as SIMOME channels are a special class of MIMOME  channels, Theorem \ref{th:esr_simome} also shows that no unique distribution is optimal for all MIMOME channel statistics. 

Before proving Theorem~\ref{th:esr_simome}, we introduce several quantities and establish properties that will be useful in the proof.

For any distribution \(d \in \{p_{(u,x)}, q_{(u,x)}\}\), we define the function \(I_d\) over \(\mathbb{R}_{\geq 0}\) as:
\begin{equation}
    \label{eq:I^d_def}
    I_d(\gamma) \triangleq I_d(u;\sqrt{\gamma}x+w),
\end{equation}
where \(w\) is a Gaussian noise with unit variance. The parameter \(\gamma\) is commonly referred to as the signal-to-noise ratio (SNR). We also denote by \(\Delta\) the difference function:
\begin{equation}
    \Delta(\gamma) \triangleq I_p(\gamma) - I_q(\gamma).
\end{equation}

The random variables $u - x - y$ form a Markov chain, so the mutual information can be decomposed as:
\begin{equation}
    \label{eq:esr_simome_I_p_decomposed}
    I_p(\gamma) = I_p(x;y) - I_p(x;y|u).
\end{equation}

From \eqref{eq:esr_simome_I_p_decomposed}, it follows that $I_p$ is infinitely differentiable, as it is the difference of two infinitely differentiable functions~\cite{GuoIMMSE, GuoMMSE}. Therefore, the function \(\Delta\) is also infinitely differentiable. Its first and second order derivatives are denoted by \(\Delta'\) and \(\Delta''\), respectively. The function satisfies the following properties, which are stated in Lemma~\ref{lem:esr_proof}.

\begin{lemma}\label{lem:esr_proof}
    \begin{equation}
        \Delta(0) = \Delta'(0) = \lim_{\gamma \rightarrow +\infty} \Delta'(\gamma)=0.
    \end{equation}
\end{lemma}

\begin{proof}[Proof of Lemma \ref{lem:esr_proof}]
    When \(\gamma = 0\), the output \(y\) is always 0, so both mutual informations $I_p$ and $I_q$ are zero. Hence, we have:
    \begin{equation}
        \Delta(0) = 0.
    \end{equation}
    The derivative \(\Delta'\) can be computed using the \textit{I-MMSE} formula~\cite{GuoIMMSE} (the factor \(\frac{1}{2}\) disappears since the random variables are complex):
    \begin{align}
        \Delta'(\gamma) &= \left(\textit{MMSE}_p(x|\gamma,y) - \textit{MMSE}_p(x|\gamma,y,u)\right)- \left(\textit{MMSE}_q(x|\gamma,y) - \textit{MMSE}_q(x|\gamma,y,u)\right).
    \end{align}
    So: 
    \begin{align}
        \Delta'(0) &= \left(\textit{MMSE}_p(x|\gamma=0) - \textit{MMSE}_p(x|u)\right)- \left(\textit{MMSE}_q(x|\gamma=0) - \textit{MMSE}_q(x|u)\right) \\
        &= \left(\textit{MMSE}_p(x|\gamma=0) - \textit{MMSE}_q(x|\gamma=0)\right) - \left(\textit{MMSE}_p(x|u) - \textit{MMSE}_q(x|u)\right) \\
        &= \left(\mathbb{E}_{x\sim p}\left[ |x|^2\right]- \mathbb{E}_{x \sim q}\left[ |x|^2\right]\right) - \left(\textit{MMSE}_c(x|u) - \textit{MMSE}_c(x|u)\right) \\
        &= 0.
    \end{align}
    Moreover, using this general bound on the \textit{MMSE} \cite{GuoMMSE}:
    \begin{equation}
        \textit{MMSE}(x|\gamma,y,u) \leq \textit{MMSE}(x|\gamma,y) \leq \frac{1}{\gamma},
    \end{equation}
    we obtain by the triangular inequality:
    \begin{align}
        \lim_{\gamma \rightarrow +\infty} |\Delta'(\gamma)| \leq    \lim_{\gamma \rightarrow +\infty} \frac{4}{\gamma} = 0,
    \end{align}
    so $\lim_{\gamma \rightarrow +\infty} \Delta'(\gamma)=0$.
\end{proof}

\begin{proof}[Proof of Theorem \ref{th:esr_simome}]
    The precoder \(c_{x|u}\) is fixed, and two distributions \(p_{(u,x)} = p_u c_{x|u}\) and \(q_{(u,x)} = q_u c_{x|u}\) are given such that \(\mathbb{E}\left[ |x|^2 \right] = \mathcal{P}_T\). We construct channel distributions \(p_{\mathbf{h}}\) and \(p_{\mathbf{g}}\) such that \eqref{eq:th_esr_trivial} or \eqref{eq:th_esr} holds.
    
    Now, assume that \(\Delta\) is concave on all of \(\mathbb{R}_{\geq 0}\). Then, \(\Delta'\) is a non-increasing function on \(\mathbb{R}_{\geq 0}\). Since \(\Delta'(0) = \Delta'(+\infty)\), it follows that \(\Delta' = 0\) and \(\Delta\) is constant. Moreover, \(\Delta(0) = 0\), so \(\Delta = 0\), which implies that for all \(\gamma \geq 0\):
    \begin{equation}
        I_p(u;\sqrt{\gamma} x+w) = I_q(u;\sqrt{\gamma} x +w).
    \end{equation}
    So for any channels statistics $p_\mathbf{h}$ and $p_\mathbf{g}$, we have:
    \begin{align}
        \textit{ESR}&(p_{(u,x)},p_\mathbf{h},p_\mathbf{g}) \nonumber\\ &= \mathbb{E}_{\mathbf{h},\mathbf{g}} \left[ I_p(u;\lVert \mathbf{h} \rVert x+w) - I_p(u;\lVert \mathbf{g} \rVert x+w)\right] \\
        &= \mathbb{E}_{\mathbf{h},\mathbf{g}}\left[ I_q(u;\lVert \mathbf{h} \rVert x+w) - I_q(u;\lVert \mathbf{g} \rVert x+w)\right] \\
        &= \textit{ESR}(q_{(u,x)},p_\mathbf{h},p_\mathbf{g}).
    \end{align}
    Thus, if \(\Delta\) is concave on all of \(\mathbb{R}_{\geq 0}\), then statement \textit{(i)} holds. We now assume that statement \textit{(i)} does not hold and proceed to prove statement \textit{(ii)}. Since both statements cannot be true for the same distributions \(p_{(u,x)}\) and \(q_{(u,x)}\), this is sufficient to show that exactly one statement is always true. 

    When \textit{(i)} does not hold, \(\Delta\) is not concave on all of \(\mathbb{R}_{\geq 0}\). So there exists a non-negative real number \(\gamma_0\) such that \(\Delta''(\gamma_0) > 0\). By continuity of $\Delta''$ and $\Delta'$, there exists an interval \(\mathcal{J} \subseteq \mathbb{R}_{\geq 0}\) over which \(\Delta\) is strictly convex and strictly monotone.

    Let \(\gamma_1 < \gamma_2\) be two numbers in \(\mathcal{J}\). We denote by \(\bar{\gamma}\) the mean of these two numbers, defined as:
    \begin{equation}
        \bar{\gamma} \triangleq \frac{\gamma_1 + \gamma_2}{2}.
    \end{equation}

    Let \(\gamma^*\) be the number in \(\mathcal{J}\) such that
    \begin{equation}
        \Delta(\gamma^*) = \frac{\Delta(\gamma_1) + \Delta(\gamma_2)}{2}.
    \end{equation}

    If \(\Delta\) is increasing over \(\mathcal{J}\), we define a channel \(\mathbf{h}_0\) such that \(\bar{\gamma} < \lVert \mathbf{h}_0 \rVert^2< \gamma^*\). If \(\Delta\) is decreasing over \(\mathcal{J}\), we define \(\mathbf{h}_0\) such that \(\bar{\gamma} < \lVert \mathbf{h}_0 \rVert^2\). These constructions for both monotonicities are illustrated in Figure~\ref{fig:Delta}.

    \begin{figure}[htbp]
        \centering

        \begin{subfigure}[t]{0.48\textwidth}
            \centering
            \includegraphics[width=\linewidth]{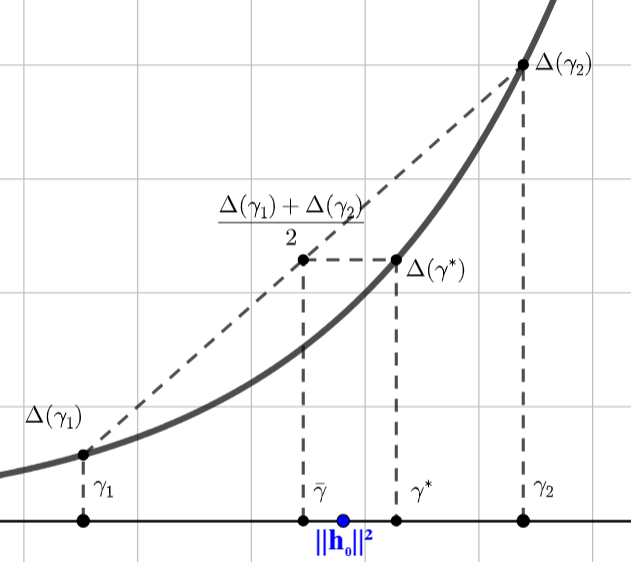}
            \caption{Increasing $\Delta$ case}\label{fig:SIMOME_Delta_preuve1}
        \end{subfigure}
        \hfill
        \begin{subfigure}[t]{0.48\textwidth}
            \centering
            \includegraphics[width=\linewidth]{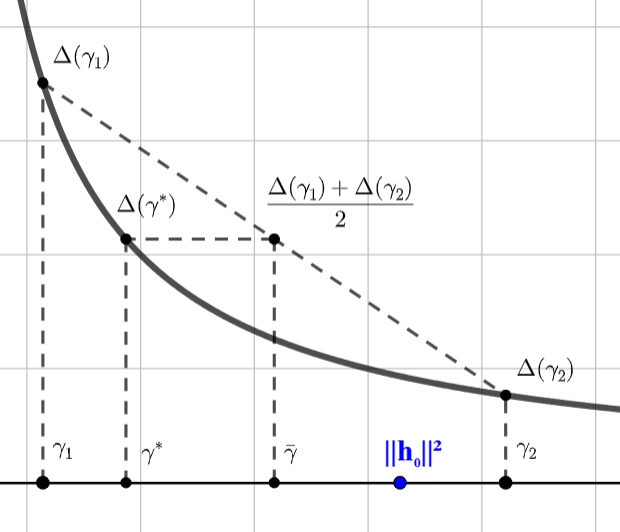}
            \caption{Decreasing $\Delta$ case}
            \label{fig:SIMOME_Delta_preuve2}
        \end{subfigure}

        \caption{Illustration of the construction of $\gamma_1$, $\gamma_2$ and $\mathbf{h}_0$.}\label{fig:Delta}
    \end{figure}

    We now construct \({\mathbf{h}}\) and \({\mathbf{g}}\) with finite support and whose square norm lies in \(\mathcal{J}\). We define the channel distribution \(p_{\mathbf{h}}\) to have a single mass point, while \(p_{\mathbf{g}}\) has two mass points of equal weights. This models a non-varying main channel and an eavesdropper channel that can take only two distinct states. We choose the two mass points of \(p_{\mathbf{g}}\) to have square norms \(\gamma_1\) and \(\gamma_2\), and the single mass point of \(p_{\mathbf{h}}\) to be \(\mathbf{h}_0\). We then have:
    \begin{align}
        &\textit{ESR}(p_{(u,x)},p_{\mathbf{h}},p_{\mathbf{g}}) - \textit{ESR}(q_{(u,x)},p_{\mathbf{h}},p_{\mathbf{g}}) \nonumber \\
        &= \mathbb{E}_{\mathbf{h},\mathbf{g}} \left[\left(I_p\left(u;\lVert\mathbf{h}\rVert x+w\right)-I_p\left(u;\lVert\mathbf{g}\rVert x+w\right)\right)\right]- \mathbb{E}_{\mathbf{h},\mathbf{g}} \left[\left(I_q\left(u;\lVert\mathbf{h}\rVert x+w\right)-I_q\left(u;\lVert\mathbf{g}\rVert x+w\right)\right)\right] \\
        &= \left( I_p(\lVert \mathbf{h}_0 \rVert^2) -\frac{I_p(\gamma_1) + I_p(\gamma_2)}{2} \right)- \left( I_q(\lVert \mathbf{h}_0 \rVert^2) -\frac{I_q(\gamma_1) + I_q(\gamma_2)}{2} \right) \\
        &= \Delta(\lVert \mathbf{h}_0 \rVert^2) - \frac{\Delta(\gamma_1) + \Delta(\gamma_2)}{2} \\
        &= \Delta(\lVert \mathbf{h}_0 \rVert^2) - \Delta(\gamma^*) <0.
    \end{align}
    So: $\textit{ESR}(p_{(u,x)},p_{\mathbf{h}},p_{\mathbf{g}}) < \textit{ESR}(q_{(u,x)},p_{\mathbf{h}},p_{\mathbf{g}})$.
\end{proof}

\begin{corollary}\label{rem:esr_simome}
    If the fixed precoder $c_{x|u}$ is such that $I_p$ is concave in the input distribution $p_x$, then both \textit{ESR} in \eqref{eq:th_esr} can be chosen to be non negative. For instance, this assumption holds if the precoder is such that $u\equiv x$ (equivalent to no precoder). 
\end{corollary}

\begin{proof}[Proof of Corollary \ref{rem:esr_simome}]
    Using the distribution $p_\mathbf{h}$ and $p_\mathbf{g}$ as defined in the proof of Theorem \ref{th:esr_simome}, we obtain the following inequalities by the concavity of $I_p$:
    \begin{align}
        \mathbb{E}_{\mathbf{h}} \left[ I_p(u;\lVert \mathbf{h} \rVert x + w)\right] &= I_p(\lVert \mathbf{h}_0 \rVert^2) \label{eq:esr_mi_concavity_step1}\\
        &\geq I_p(\bar{\gamma})\label{eq:esr_mi_concavity_step2}\\
        &\geq \frac{I_p(\gamma_1) + I_p(\gamma_2)}{2} \label{eq:esr_mi_concavity_step3}\\
        &= \mathbb{E}_{\mathbf{g}} \left[ I_p(u;\lVert \mathbf{g} \rVert x + w)\right] \label{eq:esr_mi_concavity_step4},
    \end{align}
    where \eqref{eq:esr_mi_concavity_step2} follows from the fact that \(I_p\) is non-decreasing,
    and \eqref{eq:esr_mi_concavity_step3} follows from the concavity of \(I_p\).

    So: $0 \leq \textit{ESR}(p_{(u,x)},p_{\mathbf{h}},p_{\mathbf{g}})$.
\end{proof}

\subsection{Bounds on the maximum \textit{ESR}}

In the general case, without the \textit{less noisy on average} order assumption, the maximum \textit{ESR} is difficult to derive. However, bounds can be established.

\begin{theorem}
    \label{th:esr_bounds_mimome}
    The maximum \textit{ESR} at a fixed covariance matrix $\mathbf{K}_{\mathbf{x}}$ for $\mathbf{x}$ can be bounded as follows:
    \begin{align}
        C_S^- \leq \max_{\substack{p_{u,\mathbf{x}} \\ \mathrm{cov}(\mathbf{x})=\mathbf{K}_{\mathbf{x}}}} \textit{ESR} \leq C_S^+,
    \end{align}
    with 
    \begin{align}
        C_S^- = \mathbb{E}_{\mathbf{H},\mathbf{G}}\left[ \log\left( \frac{\det(\mathbf{I}+\mathbf{H}\mathbf{K}_{\mathbf{x}}\mathbf{H}^\dagger)}{\det(\mathbf{I}+\mathbf{G}\mathbf{K}_{\mathbf{x}}\mathbf{G}^\dagger)} \right) \right] + \max_{0 \preceq \mathbf{K}_{M} \preceq \mathbf{K}_{\mathbf{x}}}  \mathbb{E}_{\mathbf{H},\mathbf{G}}\left[ \log\left( \frac{\det(\mathbf{I}+\mathbf{G}\mathbf{K}_{M}\mathbf{G}^\dagger)}{\det(\mathbf{I}+\mathbf{H}\mathbf{K}_{M}\mathbf{H}^\dagger)} \right) \right],
    \end{align}
    and
    \begin{align}
        C_S^+ = \int_{}^{} \log(\mathbf{I} + \mathbf{A} \mathbf{K}_{\mathbf{x}} \mathbf{A}^\dagger) (p_\mathbf{H}(\mathbf{A}) - p_\mathbf{G}(\mathbf{A}))^+ d\mathbf{A}.
    \end{align}
\end{theorem}

\begin{corollary}
    In Theorem \ref{th:esr_bounds_mimome}, $C_S^+$ can also be written using a conditional expectation:
    \begin{align}
        C_S^+ &= \mathbb{P} \left[ (\mathbf{H},\mathbf{G})\in \mathcal{A}^2 \right]\cdot \mathbb{E}_{\mathbf{H},\mathbf{G}}\left[ \log\left( \frac{\det(\mathbf{I}+\mathbf{H}\mathbf{K}_{\mathbf{x}}\mathbf{H}^\dagger)}{\det(\mathbf{I}+\mathbf{G}\mathbf{K}_{\mathbf{x}}\mathbf{G}^\dagger)} \right) \Big| (\mathbf{H},\mathbf{G})\in \mathcal{A}^2 \right],
    \end{align}
    with
    \begin{align}
        \mathcal{A} = \{ \mathbf{A}:p_\mathbf{G} (\mathbf{A}) \leq p_\mathbf{H} (\mathbf{A}) \}.
    \end{align}
\end{corollary}

\begin{proof}[Proof of Theorem \ref{th:esr_bounds_mimome}]
    We first prove that $C_S^- \leq \max \textit{ESR}$, then we prove that $\max \textit{ESR} \leq C_S^+$.

    We prove that $C_S^- \leq \max \textit{ESR}$ by designing an input distribution $p_{(u,\mathbf{x})}$ such that $C_S^- = \textit{ESR}(p_{(u,\mathbf{x})})$. We decompose the signal $\mathbf{x}$ into two components:
    \begin{align}
        \mathbf{x} = \mathbf{x}_I + \mathbf{x}_M,
    \end{align}
    where we choose $\mathbf{x}_I$ and $\mathbf{x}_M$ to be Gaussian and independant. The covariance of $\mathbf{x}_M$ is denoted by $\mathbf{K}_M$ so $0 \preceq \mathbf{K}_M \preceq \mathbf{K}_\mathbf{x}$. We also choose $u \equiv \mathbf{x}_I$. This is similar to an artificial noise scheme \cite{Goel} as $\mathbf{x}_I$ bears all the information and $\mathbf{x}_M$ is an additionnal Gaussian noise sometimes called a \textit{mask}. However, it is to be noted that in our proof, $\mathbf{x}_M$ is not assumed to lie in the null space of $\mathbf{H}$, as the legitimate channel is unknown to the transmitter. Therefore, the mutual information difference obtained with the defined input distribution and given instantaneous channels $\mathbf{H}$ and $\mathbf{G}$ is:
    \begin{align}
        &\log\left( \frac{\det(\mathbf{I}+\mathbf{H}\mathbf{K}_{\mathbf{x}}\mathbf{H}^\dagger)}{\det(\mathbf{I}+\mathbf{H}\mathbf{K}_{M}\mathbf{H}^\dagger)} \right) - \log\left( \frac{\det(\mathbf{I}+\mathbf{G}\mathbf{K}_{\mathbf{x}}\mathbf{G}^\dagger)}{\det(\mathbf{I}+\mathbf{G}\mathbf{K}_{M}\mathbf{G}^\dagger)} \right) \\
        &= \log\left( \frac{\det(\mathbf{I}+\mathbf{H}\mathbf{K}_{\mathbf{x}}\mathbf{H}^\dagger)}{\det(\mathbf{I}+\mathbf{G}\mathbf{K}_{\mathbf{x}}\mathbf{G}^\dagger)} \right) + \log\left( \frac{\det(\mathbf{I}+\mathbf{G}\mathbf{K}_{M}\mathbf{G}^\dagger)}{\det(\mathbf{I}+\mathbf{H}\mathbf{K}_{M}\mathbf{H}^\dagger)} \right).
    \end{align}

    Averaging over all instantaneous channels $\mathbf{H}$ and $\mathbf{G}$ and maximizing over all possible $\mathbf{K}_M$ yields $C_S^- \leq \max \textit{ESR}$.

    Now, we upperbound $\max \textit{ESR}$ to show that $\max \textit{ESR} \leq C_S^+$. We have:
    \begin{align}
        \max \textit{ESR} &= \max \mathbb{E}_{\mathbf{H},\mathbf{G}} \left[ I(u;\mathbf{y}) - I(u;\mathbf{z}) \right] \\
        &= \max \mathbb{E}_{\mathbf{H}} \left[ I(u;\mathbf{y})\right] -  \mathbb{E}_{\mathbf{G}} \left[I(u;\mathbf{z}) \right] \\
        &= \max \int I(u;\mathbf{y}|\mathbf{H}=\mathbf{A}) p_\mathbf{H}(\mathbf{A}) d\mathbf{A} - \int I(u;\mathbf{z}|\mathbf{G}=\mathbf{A}) p_\mathbf{G}(\mathbf{A}) d\mathbf{A} \\
        &= \max \int I(u;\mathbf{y}|\mathbf{H}=\mathbf{A}) (p_\mathbf{H}(\mathbf{A})-p_\mathbf{G}(\mathbf{A})) d\mathbf{A} \label{eq:esr_bounds_mimome_proof2_step4}\\
        &\leq \int \max I(u;\mathbf{y}|\mathbf{H}=\mathbf{A}) (p_\mathbf{H}(\mathbf{A})-p_\mathbf{G}(\mathbf{A}))^+ d\mathbf{A} \label{eq:esr_bounds_mimome_proof2_step5}\\
        &= \int \log(\det(\mathbf{I} + \mathbf{A} \mathbf{K}_\mathbf{x}\mathbf{A})) (p_\mathbf{H}(\mathbf{A})-p_\mathbf{G}(\mathbf{A}))^+ d\mathbf{A}, \label{eq:esr_bounds_mimome_proof2_step6}
    \end{align}
    where \eqref{eq:esr_bounds_mimome_proof2_step4} follows from the fact that $\mathbf{y}$ and $\mathbf{z}$ have the same distribution when $\mathbf{H}=\mathbf{G}$, \eqref{eq:esr_bounds_mimome_proof2_step5} follows from the non negativity of mutual information and \eqref{eq:esr_bounds_mimome_proof2_step6} follows from the maximization of mutual information with a fixed covariance matrix. 
\end{proof}

The lower bound in Theorem \ref{th:esr_bounds_mimome} can be detailled and the upperbound can be refined in a SIMOME scenario.

\begin{theorem}
    \label{th:esr_bounds_simome}
    If $N_A=1$ (SIMOME scenario), the maximum \textit{ESR} can be bounded:
    \begin{align}
        C_{S,\text{SIMOME}}^- \leq \max_{\substack{p_{u,x} \\  \mathbb{E}[|x|^2] \leq \mathcal{P}_T}} \textit{ESR} \leq C_{S,\text{SIMOME}}^+,
    \end{align}
    with 
    \begin{align}
        C_{S,\text{SIMOME}}^- = \max_{0 \leq P \leq \mathcal{P}_T} \mathbb{E}_{\mathbf{h},\mathbf{g}} \left[ \log \left( \frac{1+ \mathcal{P}_T - P + P \lVert \mathbf{h} \rVert^2}{1+ \mathcal{P}_T - P + P \lVert \mathbf{g} \rVert^2} \right) \right],
    \end{align}
    and 
    \begin{align}
        C_{S,\text{SIMOME}}^{+} = \min\{C_{S,\text{SIMOME}}^{+(0)}, C_{S,\text{SIMOME}}^{+(1)}\},
    \end{align}
    where
    \begin{align}
        &C_{S,\text{SIMOME}}^{+(0)} = \int_{\mathbb{R}\geq 0} \log(1+\mathcal{P}_T a^2) (p_{\lVert \mathbf{h} \rVert}(a)-p_{\lVert \mathbf{g} \rVert}(a))^+ da \\
        &C_{S,\text{SIMOME}}^{+(1)} = \int_{\mathbb{R}\geq 0} \frac{\mathcal{P}_T(\mathbb{P}\left[\lVert \mathbf{h} \rVert\geq a \right]-\mathbb{P}\left[\lVert \mathbf{g} \rVert\geq a \right])^+}{1+\mathcal{P}_T a^2} da \\
    \end{align}
\end{theorem}

\begin{corollary}
    In Theorem \ref{th:esr_bounds_simome}, $C_{S,\text{SIMOME}}^{+(0)}$ and $C_{S,\text{SIMOME}}^{+(1)}$ can also be written using a conditional expectation:
    \begin{align}
        C_{S,\text{SIMOME}}^{+(0)} &= \mathbb{P} \left[ (\mathbf{h},\mathbf{g})\in \mathcal{A}_{(0)}^2 \right] \cdot \mathbb{E}_{\mathbf{h},\mathbf{g}}\left[ \log\left( \frac{1+\lVert \mathbf{h} \rVert^2 \mathcal{P}_T}{1+\lVert \mathbf{g} \rVert^2 \mathcal{P}_T} \right) \Big| (\lVert \mathbf{h} \rVert,\lVert \mathbf{g} \rVert)\in \mathcal{A}_{(0)}^2 \right],
    \end{align}
    \begin{align}
        C_{S,\text{SIMOME}}^{+(1)} &= \mathbb{P} \left[ (\mathbf{h},\mathbf{g})\in \mathcal{A}_{(1)}^2 \right] \cdot \mathbb{E}_{\mathbf{h},\mathbf{g}}\left[ \log\left( \frac{1+\lVert \mathbf{h} \rVert^2 \mathcal{P}_T}{1+\lVert \mathbf{g} \rVert^2 \mathcal{P}_T} \right) \Big| (\lVert \mathbf{h} \rVert,\lVert \mathbf{g} \rVert)\in \mathcal{A}_{(1)}^2 \right],
    \end{align}
    with 
    \begin{align}
        &\mathcal{A}_{(0)} = \{a: p_{\lVert \mathbf{g} \rVert}(a)\leq  p_{\lVert \mathbf{h} \rVert}(a)\}, \\
        &\mathcal{A}_{(1)} = \{a: \mathbb{P}[\lVert \mathbf{g} \rVert \geq a] \leq \mathbb{P}[\lVert \mathbf{h} \rVert \geq a] \}.
    \end{align}
\end{corollary}

\begin{proof}[Proof of Theorem \ref{th:esr_bounds_simome}]
    The lower bound $C_{S,\text{SIMOME}}^-$ and the upper bound $C_{S,\text{SIMOME}}^{+(0)}$ are obtained applying Theorem \ref{th:esr_bounds_mimome} to the SIMOME scenario. The upper bound $C_{S,\text{SIMOME}}^{+(1)}$ however is specific to the SIMOME case. We have:
    \begin{align}
        \max \textit{ESR}
        &= \max \mathbb{E}_{\mathbf{H},\mathbf{G}} \left[ I(u;\mathbf{y}) - I(u;\mathbf{z}) \right] \\
        &= \max \mathbb{E}_{\mathbf{H}} \left[ I(u;\mathbf{y})\right] -  \mathbb{E}_{\mathbf{G}} \left[I(u;\mathbf{z}) \right] \\
        &= \max \int \frac{d}{d(a^2)}\left(I(u;\mathbf{y}|\lVert \mathbf{h} \rVert=a)\right) \mathbb{P}[\lVert \mathbf{h} \rVert \geq a] da - \int \frac{d}{d(a^2)}\left(I(u;\mathbf{z}|\lVert \mathbf{g} \rVert=a)\right) \mathbb{P}[\lVert \mathbf{g} \rVert \geq a] da \\
        &= \max \int \frac{d}{d(a^2)}\left(I(u;\mathbf{y}|\lVert \mathbf{h} \rVert=a)\right) \cdot(\mathbb{P}[\lVert \mathbf{h} \rVert \geq a] - \mathbb{P}[\lVert \mathbf{g} \rVert \geq a]) da \label{eq:esr_bounds_simome_proof1_step4}\\
        &\leq \int \max \frac{d}{d(a^2)}\left(I(u;\mathbf{y}|\lVert \mathbf{h} \rVert=a)\right) \cdot(\mathbb{P}[\lVert \mathbf{h} \rVert \geq a] - \mathbb{P}[\lVert \mathbf{g} \rVert \geq a])^+ da \label{eq:esr_bounds_simome_proof1_step5}
    \end{align}
    where \eqref{eq:esr_bounds_simome_proof1_step4} comes from the fact that $\mathbf{y}$ and $\mathbf{z}$ have the same distribution when $\mathbf{h}=\mathbf{g}$ and \eqref{eq:esr_bounds_simome_proof1_step5} comes from the non negativity of mutual information. 

    We now derive the maximum of the derivative of the mutual information. Note that this derivative cannot be expressed as a single \textit{MMSE}, as in~\cite{GuoIMMSE}, because the input $u$ is an auxiliary variable and cannot be directly identified with $x$. Instead, we express the derivative as the difference of two mutual informations evaluated at nearby SNR values.
    \begin{align}
        \frac{d}{d(a^2)}\left(I(u;\mathbf{y}|\lVert \mathbf{h} \rVert=a)\right) &= \frac{1}{2a} \frac{d}{da}\left(I(u;\mathbf{y}|\lVert \mathbf{h} \rVert=a)\right)\\
        &= \frac{1}{2a} \lim_{\epsilon \rightarrow 0} \frac{I(u;\mathbf{y}|\lVert \mathbf{h} \rVert=a+\epsilon)-I(u;\mathbf{y}|\lVert \mathbf{h} \rVert=a)}{\epsilon}
    \end{align}

    However, for every $\epsilon\geq 0$, we know by \cite{Csiszar} that the maximum difference between those two mutual informations is the secrecy capacity of the induced wiretap channel. Thus, by \cite{SLYC}, we have:
    \begin{align}
        \max_{p_{(u,x)}} \left[ I(u;\mathbf{y}|\lVert \mathbf{h} \rVert=a+\epsilon)-I(u;\mathbf{y}|\lVert \mathbf{h} \rVert=a) \right] = \log\left( \frac{1+ \mathcal{P}_T(a+\epsilon)^2}{1+ \mathcal{P}_Ta^2} \right).
    \end{align}
    We obtain:
    \begin{align}
        \label{eq:esr_bounds_simome_proof4}
        \max \frac{d}{d(a^2)}&\left(I(u;\mathbf{y}|\lVert \mathbf{h} \rVert=a)\right) = \frac{\mathcal{P}_T}{1+\mathcal{P}_Ta^2}
    \end{align}
    Finally, plotting \eqref{eq:esr_bounds_simome_proof4} into \eqref{eq:esr_bounds_simome_proof1_step5} concludes the proof. 
\end{proof}

%%%%%%%%%%%%%%%%%%%%%%%%%%%%%%%%%%%%%%%%%%%%%%%%%%%
%%                                               %%
%%               NOUVELLE SECTION                %%
%%                                               %%
%%%%%%%%%%%%%%%%%%%%%%%%%%%%%%%%%%%%%%%%%%%%%%%%%%%

\section{Rayleigh fading channels}
\label{sec:rayleigh}

\subsection{Degraded MIMOME Rayleigh fading channels}

In this subsection, we consider the special case of degraded MIMOME Rayleigh fading channels. The legitimate channel is Rayleigh distributed, so each coefficient of the matrix follows $\mathcal{N}(0, \sigma_{\mathbf{H}}^2)$. The eavesdropper channel is $\mathbf{G}$ and can be expressed as a cascade of two Rayleigh channels: the first one is $\mathbf{H}$ and the second one is $\Tilde{\mathbf{G}}$, whose coefficient follows $\mathcal{N}(0, \Tilde{\sigma_{\mathbf{G}}}^2)$. As the eavesdropper channel is degraded compared to the legitimate channel, channel $\mathbf{H}$ is also less noisy than channel $\mathbf{G}$. Therefore, $\mathbf{G}=\mathbf{H} \Tilde{\mathbf{G}}$. The \textit{SOP}, the \textit{ESR} and the \textit{EPSR} can be computed in this case.

From Theorems \ref{th:sop_mimome}, \ref{th:epsr_mimome} and \ref{th:esr_mimome}, the \textit{SOP}, the \textit{EPSR} and the \textit{ESR} are optimized by the Gaussian input when the input covariance $\mathbf{K}_\mathbf{x}$ is fixed. We can now derive the input covariance matrix that maximizes the secrecy metrics.

First, we derive the maximum \textit{ESR} and \textit{EPSR}. For all fixed instantaneous channel state $(\mathbf{H},\mathbf{G})$, $\mathbf{x}-\mathbf{y}-\mathbf{z}$ forms a Markov chain and:
\begin{align}
    \label{eq:rayleigh_degraded}
    I(\mathbf{x};\mathbf{y}) - I(\mathbf{x};\mathbf{z}) = I(\mathbf{x};\mathbf{y}|\mathbf{z}).
\end{align}

The mutual information is always non negative, so in this case, the \textit{EPSR} is equal to the \textit{ESR} as the term in the expectancy is never forced to zero by the positive part operator. From \cite{Khisti}, we know that when $\mathbf{x}$ is Gaussian, the right hand side of \eqref{eq:rayleigh_degraded} is concave in the covariance matrix $\mathbf{K}_\mathbf{x}$. Thus the function
\begin{align}
    \label{eq:rayleigh_degraded_logdet}
    \mathbf{K}_\mathbf{x} \mapsto \log \left( \frac{\det (\mathbf{I} + \mathbf{H}^\dagger \mathbf{K}_\mathbf{x} \mathbf{H})}{\det (\mathbf{I} + \mathbf{G}^\dagger \mathbf{K}_\mathbf{x} \mathbf{G})} \right)
\end{align}
is concave in $\mathbf{K}_\mathbf{x}$ and the average over all instantaneous channel state $(\mathbf{H},\mathbf{G})$ is also concave in $\mathbf{K}_\mathbf{x}$. This means that the \textit{ESR} is concave in $\mathbf{K}_\mathbf{x}$.

From \cite{Telatar}, we know that the covariance matrix $\mathbf{K}_\mathbf{x} = \frac{\mathcal{P}_T}{N_A}\mathbf{I}$ maximizes both $\mathbb{E}\left[ \log(\det(\mathbf{I} + \mathbf{H}^\dagger \mathbf{K}_\mathbf{x} \mathbf{H}))\right]$ and $\mathbb{E}\left[ \log(\det(\mathbf{I} + \mathbf{G}^\dagger \mathbf{K}_\mathbf{x} \mathbf{G}))\right]$. Thus, this input is a stationnary point of the \textit{ESR}. As the \textit{ESR} is a concave function, the maximum is achieved by this input. This leads to
\begin{align}
    \label{eq:rayleigh_degraded_esr_epsr}
    \textit{ESR}^\star =\textit{EPSR}^\star = \mathbb{E}_{\mathbf{H},\mathbf{G}} \left[\log\left( \frac{\det (\mathbf{I} + \frac{\mathcal{P}_T}{N_A} \mathbf{H}^\dagger \mathbf{H})}{\det (\mathbf{I} + \frac{\mathcal{P}_T}{N_A} \mathbf{G}^\dagger \mathbf{G})} \right)\right].
\end{align}

We now optimize the \textit{SOP} and show that the covariance matrix $\mathbf{K}_\mathbf{x} = \frac{\mathcal{P}_T}{N_A}\mathbf{I}$ also minimizes the \textit{SOP}.

We apply the eigenvalue decomposition to $\mathbf{K}_\mathbf{x}$, and we obtain $\mathbf{K}_\mathbf{x}=\mathbf{U} \mathbf{D} \mathbf{U}^\dagger$ for some unitary matrix $\mathbf{U}$ and some diagonal matrix $\mathbf{D}$. As the channels are Rayleigh distributed, the distribution of the secrecy rate is the same whether $\mathbf{K}_\mathbf{x}=\mathbf{U} \mathbf{D} \mathbf{U}^\dagger$ or $\mathbf{K}_\mathbf{x}=\mathbf{D}$. Thus, we can restrict the search of an optimal covariance matrix to diagonal covariance matrices. 

Let $D_1 \triangleq Diag(d_1,...,d_{N_A})$ be an optimal covariance matrix whose trace is equal to the average power contraint $\mathcal{P}_T$. As the channels are Rayleigh faded, no direction preveils on another, so for any $i\in \{1,...,N_A\}$, $D_i \triangleq Diag(d_i,d_i+1,...,d_{N_A},d_1,...,d_{i-1})$ is also optimal. Let $p_{\text{g}}^{(i)}$ denote the Gaussian input with covariance matrix $D_i$. We construct a new distribution $\Tilde{p}$ by time sharing: the time is equally shared among the inputs $p_g^{(i)}$. We have
\begin{align}
    \textit{SOP}(\Tilde{p}) = \sum_{i=1}^{N_A} \frac{\textit{SOP}\left(p_{\text{g}}^{(i)}\right)}{N_A} = \textit{SOP}\left(p_{\text{g}}^{(1)}\right).
\end{align}
The covariance matrix of the input $\Tilde{p}$ is the average of all $D_i$'s, which is $\frac{\mathcal{P}_T}{N_A}\mathbf{I}$. Therefore, by Theorem \ref{th:sop_mimome}, $\textit{SOP}(\Tilde{p}) \geq \textit{SOP}(p^\star)$, where $p^\star$ is the Gaussian input with covariance $\frac{\mathcal{P}_T}{N_A}\mathbf{I}$. As $\textit{SOP}\left(\Tilde{p}\right) = \textit{SOP}(p_{\text{g}}^{(1)})$ and $p_{\text{g}}^{(1)}$ is minimal, $p^\star$ is also minimal. We have:

\begin{align}
    \label{eq:rayleigh_degraded_sop}
    \textit{SOP}^\star =  \mathbb{P} \left[\log\left( \frac{\det (\mathbf{I} + \frac{\mathcal{P}_T}{N_A} \mathbf{H}^\dagger \mathbf{H})}{\det (\mathbf{I} + \frac{\mathcal{P}_T}{N_A} \mathbf{G}^\dagger \mathbf{G})} \right) < r\right].
\end{align}

\subsection{Less noisy on average MIMOME Rayleigh fading channels}

In this subsection, we study the special case of \textit{less noisy on average} MIMOME Rayleigh fading channels. The legitimate channel $\mathbf{H}$ and the eavesdropper channel $\mathbf{G}$ are Rayleigh distributed, so each coefficient of the matrix follows respectiveley $\mathcal{N}(0, \sigma_{\mathbf{H}}^2)$ and $\mathcal{N}(0, \sigma_{\mathbf{G}}^2)$. The order assumption imposes that for every input $u \rightarrow \mathbf{x}$: $\mathbb{E}\left[I(u;\mathbf{y})-I(u;\mathbf{z})\right]\geq 0$. This case happends with different set of parameters $(N_A,N_B,N_E,\sigma_{\mathbf{H}}, \sigma_{\mathbf{G}})$. For instance, when $N_B\geq N_E$ and $\sigma_{\mathbf{H}}\leq \sigma_{\mathbf{G}}$, the eavesdropper channel is equivalent to a degraded version of the legitimate channel, as shown in \cite{Lin}. The channel order also holds when $N_A=1$ and the complementary distributive function (CCDF) of $\lVert \mathbf{h} \rVert$ is always greater than the CCDF of $\lVert \mathbf{g} \rVert$, as the channel can then be expressed with equivalent degraded SISO channels. 

The \textit{less noisy on average} order is more general than the \textit{degraded} order studied in the previous subsection. However, we show that the optimal \textit{ESR} has the same mathematical expression as in the more specific case. 

Assuming the \textit{less noisy on average} channel order, we compute the maximum \textit{ESR} using a similar approach to the optimal \textit{SOP} in the degraded Rayleigh fading case. Note that the approach that was used for the maximum \textit{ESR} in the degraded case cannot be replicated as it requires to write the secrecy rate as a single mutual information that is concave in the Gaussian input covariance matrix. 

By Theorem \ref{th:esr_mimome}, the maximum \textit{ESR} is achieved by a Gaussian input with covariance $\mathbf{K}_\mathbf{x}$. We now search the $\mathbf{K}_\mathbf{x}$ that maximizes the \textit{ESR}. 

As detailled in the previous subsection, we can restrict our search to diagonal matrices. Let $D_1 \triangleq Diag(d_1,...,d_{N_A})$ be a covariance matrix whose trace is equal to $\mathcal{P}_T$ and which maximizes the \textit{ESR}. As in the previous subsection, construct $D_i$ and $p_{\text{g}}^{(i)}$. As the channels are Rayleigh faded, no direction preveils on another, so $D_i$ also maximizes the \textit{ESR}. We construct $\Tilde{p}$ by time sharing: the time is equally shared among the inputs $p_i$. We have
\begin{align}
    \textit{ESR}(\Tilde{p}) = \sum_{i=1}^{N_A} \frac{\textit{ESR}\left(p_{\text{g}}^{(i)}\right)}{N_A} = \textit{ESR}\left(p_{\text{g}}^{(1)}\right).
\end{align}
The covariance matrix of the input $\Tilde{p}$ is the average of all $D_i$'s, which is $\frac{\mathcal{P}_T}{N_A}\mathbf{I}$. Therefore, by Theorem \ref{th:esr_mimome}, $\textit{ESR}(\Tilde{p}) \leq \textit{ESR}(p^\star)$, where $p^\star$ is the Gaussian input with covariance $\frac{\mathcal{P}_T}{N_A}\mathbf{I}$. As $\textit{ESR}\left(\Tilde{p}\right) = \textit{ESR}(p_{\text{g}}^{(1)})$ and $p_{\text{g}}^{(1)}$ is optimal, $p^\star$ is also optimal. We have for less noisy on average MIMOME Rayleigh channels:
\begin{align}
    \label{eq:rayleigh_mnoa_esr_epsr}
    \textit{ESR}^\star =\textit{EPSR}^\star = \mathbb{E}_{\mathbf{H},\mathbf{G}} \left[\log\left( \frac{\det (\mathbf{I} + \frac{\mathcal{P}_T}{N_A} \mathbf{H}^\dagger \mathbf{H})}{\det (\mathbf{I} + \frac{\mathcal{P}_T}{N_A} \mathbf{G}^\dagger \mathbf{G})} \right)\right].
\end{align}

\subsection{SIMOME Rayleigh fading channels}

In this subsection, we derive closed form expression of the optimal \textit{SOP} and the optimal \textit{EPSR} in a specific SIMOME setting. We assume that the main channel $\mathbf{h}$ is constant over time and known to the transmitter, while the eavesdropper channel $\mathbf{g}$ consists of \(N_E\) independent Gaussian random variables with variance \(\sigma^2\). 
This models perfect main channel estimation at the transmitter and an unknown eavesdropper position in a rich-scattering environment.

In the described scenario, \(2 \lVert g \rVert ^2 / \sigma^2\) follows a \(\chi^2\) distribution with \(2 N_E\) degrees of freedom. By Theorem \ref{th:sop_simome}, the minimal \textit{SOP} is:
\begin{align}
    &\textit{SOP}^{\star}(r) \nonumber \\
    &= \mathbb{P}\left[ 
        \frac{1}{2^{r} \mathcal{P}_T} 
        + \frac{\lVert h \rVert ^2}{2^{r}} 
        - \frac{1}{\mathcal{P}_T} 
        \leq \lVert g \rVert ^2
    \right] \\
    &= 
    \begin{cases}
        \displaystyle 
        1- \frac{1}{(N_E-1)!}\,
        \Gamma_i\!\left(
            N_E, 
            \frac{1}{2^{r}\mathcal{P}_T \sigma^2} 
            + \frac{\lVert h \rVert ^2}{2^{r}\sigma^2} 
            - \frac{1}{\mathcal{P}_T\sigma^2}
        \right), &\text{if } r \leq \log\!\big(1+ \mathcal{P}_T \lVert h \rVert ^2\big), \\[8pt]
        1, &\text{if } r > \log\!\big(1+ \mathcal{P}_T \lVert h \rVert ^2\big),
    \end{cases}
\end{align}
where $\Gamma_i$ is the incomplete gamma function defined as:
\begin{equation}
    \Gamma_i(s,x) \triangleq \int_{0}^{x} t^{s-1}e^{-t} dt.
\end{equation}

Figure~\ref{fig:SOP_example} shows the optimal \textit{SOP} as a function of \(r\) for \(N_E = 2\), \(\mathcal{P}_T = 3\), \(\lVert h \rVert  = 5\), and different values of \(\sigma^2\). It can be observed that as the eavesdropper channel gain increases, the \textit{SOP} increases faster. When \(r\) exceeds the main channel capacity, the information cannot be transmitted with perfect secrecy, as it cannot even be reliably received by the legitimate receiver; consequently, the \textit{SOP} equals 1.

\begin{figure}[h]
    \centering
    \includegraphics[width=\linewidth]{img/courbes_SOP_exemples.png}
    \caption{\textit{SOP}$^{\star}$ as a function of $r$ with $N_A=1$, $N_E=2$, $\mathcal{P}_T=3$, $\lVert h \rVert =5$ and different values of $\sigma$.}
    \label{fig:SOP_example}
\end{figure}

Next, the optimal \textit{EPSR} can also be computed as follows:
\begin{align}
    \textit{EPSR}^\star &= \mathbb{E}\left[ \log\left( \frac{1+\mathcal{P}_T \lVert h \rVert^2}{1+ \mathcal{P}_T \lVert g \rVert^2} \right) \right] \\
    &= \int_{r=0}^{+\infty} \mathbb{P} \left( \log\left( \frac{1+\mathcal{P}_T \lVert h \rVert^2}{1+ \mathcal{P}_T \lVert g \rVert^2} \right) >r \right) dr \\
    &= \int_{r=0}^{+\infty} \left( 1- \textit{SOP}^{\star}(r)\right) dr \\
    &= \int_{r=0}^{\log(1+\mathcal{P}_T \lVert h \rVert^2)} \frac{\Gamma_i\!\left(N_E, \frac{1}{2^{r}\mathcal{P}_T \sigma} + \frac{\lVert h \rVert^2}{2^{r}\sigma} - \frac{1}{\mathcal{P}_T\sigma} \right)}{(N_E-1)!} dr.
\end{align}

%%%%%%%%%%%%%%%%%%%%%%%%%%%%%%%%%%%%%%%%%%%%%%%%%%%
%%                                               %%
%%               NOUVELLE SECTION                %%
%%                                               %%
%%%%%%%%%%%%%%%%%%%%%%%%%%%%%%%%%%%%%%%%%%%%%%%%%%%
\section{Examples and counterexamples}
\label{sec:examples}

\subsection{Illustration of Theorem \ref{th:esr_simome}}
\label{subsec:esr_example}

In this subsection, we present two pairs of channel statistics illustrating Theorem \ref{th:esr_simome}. In one pair, the non-precoded Gaussian input outperforms the non-precoded BPSK input with respect to the \textit{ESR} metric. In the other example, the non-precoded BPSK input outperforms the non-precoded Gaussian input with regard to the \textit{ESR} metric.

Both examples are based on the same model: a SISOSE wiretap channel in which the main channel fading coefficient \(h\) is constant and equal to a positive number \(h_0\), while the eavesdropper channel fading coefficient \(g\) is 0 with probability \(1/2\) and equal to a positive number \(g_0\) with probability \(1/2\). The total input power and the noise power are both normalized to 1.

The amount of information received when the input is BPSK and the SNR is \(\gamma\) is given by (e.g., \cite[Problem 4.22]{Gallager}):
\begin{align}
    I_{\text{BPSK}}(\gamma) = I_{\text{BPSK}}(x;\sqrt{\gamma} x+w) = 2\gamma\log(e) - \int_{-\infty}^{+\infty} \frac{e^{-\frac{t^2}{2}}}{\sqrt{2\pi}} \log(\cosh(2\gamma - \sqrt{2\gamma t})) dt.
\end{align}

The $I_{\text{BPSK}}$ function tends to $1$ as $\gamma$ tends to $+\infty$, so we can choose the main channel SNR $h_0$ such that: $I_{\text{BPSK}}(h_0)>\frac{3}{4}$.

\begin{itemize}
    \item \textbf{First example} (the Gaussian input outperforms the BPSK input): Choose \(g_0\) such that \(g_0 < h_0\). Then, by the DPI, any input distribution yields a positive secrecy rate. In this case, the \textit{ESR} and the \textit{EPSR} are equal. Therefore, by Theorem~\ref{th:epsr_simome}, the input distribution that maximizes the \textit{ESR} is the non-precoded Gaussian input. Specifically, the Gaussian input achieves a higher \textit{ESR} than the BPSK input.

    \item \textbf{Second example} (the BPSK input outperforms the Gaussian input): Choose $g_0$ such that:
    \begin{equation}
        \left(\frac{1+h_0}{2^{\frac{1}{4}}}\right)^2-1 <g_0<(1+h_0)^2-1.
    \end{equation} 
    Then, a Gaussian input yields a positive \textit{ESR} that is less than \(\frac{1}{4}\), as:
    \begin{equation}
        \textit{ESR}_{\text{g}} = \log(1+h_0) - \frac{1}{2}\log(1+g_0) \in \left( 0, \frac{1}{4}\right).
    \end{equation}
    However, a BPSK input yields an \textit{ESR} that is greater than \(\frac{1}{4}\), as:
    \begin{align}
        \textit{ESR}_{\text{BPSK}} &= I_{\text{BPSK}}(h_0) - \frac{1}{2}I_{\text{BPSK}}(g_0) \\
        &> I_{\text{BPSK}}(h_0) - \frac{1}{2} > \frac{1}{4}.
    \end{align}
    So, in this case, the BPSK input achieves a higher \textit{ESR} than the Gaussian input, while the Gaussian input still results in a positive \textit{ESR}.
\end{itemize}

\begin{figure}[h]
    \centering
    \includegraphics[width=\linewidth]{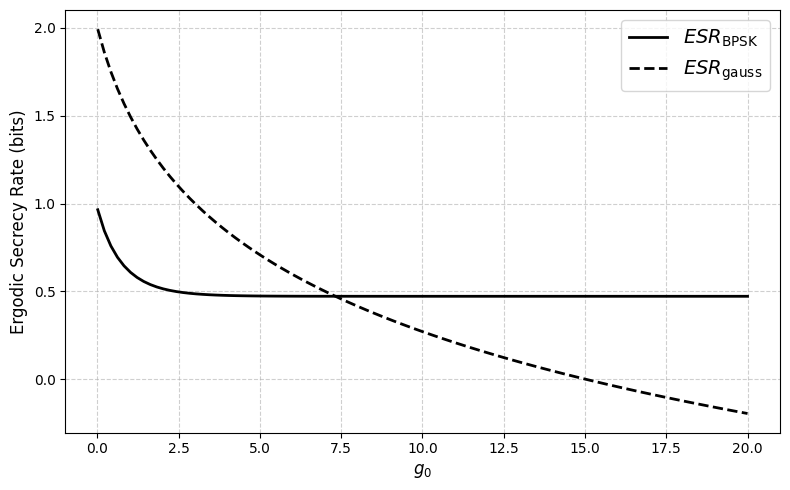}
    \caption{\textit{ESR} as a function of $g_0$ with a BPSK input and a gaussian input, with $h_0=3$.}
    \label{fig:ESR_example}
\end{figure}

Figure~\ref{fig:ESR_example} shows the \textit{ESR} for a Gaussian input and a BPSK input as functions of \(g_0\), with \(h_0 = 3\) (since \(I_{\text{BPSK}}(3) \simeq 0.97 > 0.75\)). In this case, \(\left(\frac{1 + h_0}{2^{\frac{1}{4}}}\right)^2 - 1 \simeq 10.3\) and \((1 + h_0)^2 - 1 = 15\).

\subsection{On the Importance of Channel Ordering Assumptions}
\label{subsec:degenerated}

In this subsection, we present one last scenario in which the optimal secrecy performance is not achieved by a non-precoded Gaussian input. This example demonstrates that the \textit{uniformly less noisy} assumption in Theorem~\ref{th:sop_mimome} and the \textit{less noisy on average} assumption in Theorem~\ref{th:esr_bounds_mimome} are essential and cannot be dismissed in general. %We also note that the example provided in Section~\ref{sec:esr_example} already serves as a counterexample showing that the \textit{uniformly less noisy} assumption in Theorem~\ref{th:epsr_mimome} is important.

\subsubsection{Channel distributions}
Let $N_A = 2$ and $N_B = N_E = 1$. We construct two channels, denoted by $\mathbf{h}$ and $\mathbf{g}$. The main channel $\mathbf{h}$ is assumed to be deterministic and constant over time:
\begin{equation}
    \mathbf{h} = \begin{pmatrix} 1 & 0 \end{pmatrix}.
\end{equation}

The eavesdropper channel $\mathbf{g}$ consists of one fixed component and one component with a random sign. Specifically, $\mathbf{g}$ takes two possible values, $\mathbf{g}_1$ and $\mathbf{g}_2$, each with probability $0.5$:
\begin{align}
    \mathbf{g}_1 &= \begin{pmatrix} +1 & 1 \end{pmatrix}, \\
    \mathbf{g}_2 &= \begin{pmatrix} -1 & 1 \end{pmatrix}.
\end{align}

\begin{figure}
    \centering
    \includegraphics[width=0.7\linewidth]{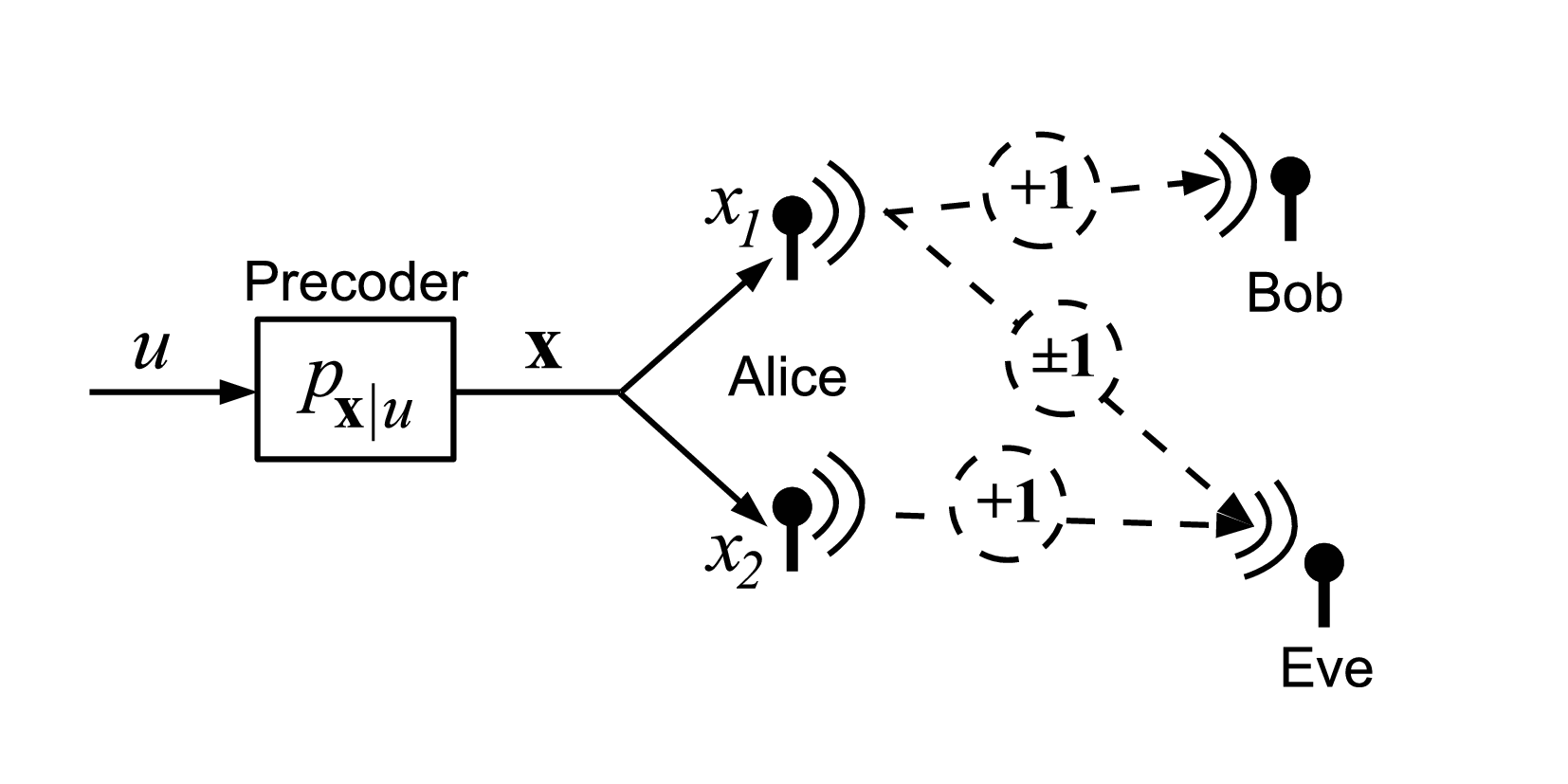}
    \caption{Counterexample illustration}
    \label{fig:counterexample}
\end{figure}

This channel model is illustrated in Fig.~\ref{fig:counterexample} The total average transmit power is constrained to $\mathcal{P}_T$.

\subsubsection{Non-precoded Gaussian input}
We first consider a Gaussian input without precoding. Its covariance matrix can be written as
\begin{align}
    \mathbf{K} = 
    \begin{pmatrix}
    a & b \\
    b^\ast & c
    \end{pmatrix},
\end{align}
where $a,c \in \mathbb{R}$ and $b \in \mathbb{C}$. By Sylvester’s criterion, the matrix $\mathbf{K}$ must satisfy $a \geq 0$, $c \geq 0$, and $ac \geq |b|^2$. In addition, the power constraint imposes $a + c = \mathcal{P}_T$. We denote by $\beta = \Re(b)$ the real part of $b$.

When the eavesdropper channel realization is $\mathbf{g}_1$, the secrecy rate is
\begin{equation}
    R_s^{(1)} = \log(1+a) - \log(1+\mathcal{P}_T + 2\beta),
\end{equation}
while for $\mathbf{g}_2$, it becomes
\begin{equation}
    R_s^{(2)} = \log(1+a) - \log(1+\mathcal{P}_T - 2\beta).
\end{equation}

The quantity $\min(R_s^{(1)}, R_s^{(2)})$ is maximized when $\beta = 0$, in which case both secrecy rates coincide and reduce to
\begin{equation}
     R_s = \log(1+a) - \log(1+\mathcal{P}_T) \leq 0.
\end{equation}

This implies that no non-precoded Gaussian input can achieve a zero \textit{SOP}, and in fact the \textit{SOP} is at least $\frac{1}{2}$.

Moreover, the \textit{ESR} is given by
\begin{align}
    \textit{ESR} &= \frac{1}{2} \left( R_s^{(1)} + R_s^{(2)} \right) \\
    &= \log(1+a) - \frac{1}{2} \log\left((1+\mathcal{P}_T)^2 - 4\beta^2 \right) \\
    &\leq \log(1+a) - \log(1+\mathcal{P}_T) \\
    &\leq 0.
\end{align}

Thus, the \textit{ESR} is also non-positive for any such Gaussian input.

\subsubsection{Gaussian signaling with artificial noise}
We now consider an alternative input strategy. Let $x_1 \equiv u$, while $x_2$ is independent of $u$ and Gaussian. Both $x_1$ and $x_2$ are allocated equal power $\mathcal{P}_T/2$.

In this case, the secrecy rate becomes independent of the eavesdropper channel realization ($\mathbf{g}_1$ or $\mathbf{g}_2$) and is given by
\begin{align}
    R_S &= \log \left(1+\frac{\mathcal{P}_T}{2} \right) 
    - \log \left(1+\frac{\frac{\mathcal{P}_T}{2}}{1+\frac{\mathcal{P}_T}{2}} \right) \\
    &= \log \left( 1+ \frac{\mathcal{P}_T^2}{4 + 4\mathcal{P}_T} \right).
\end{align}

Since $R_S > 0$, the \textit{SOP} is equal to zero for sufficiently small rates $r$. In addition, the secrecy rate coincides with the \textit{ESR}, which is therefore positive.

This example shows that, by using a more structured input (combining information-bearing signaling and artificial noise), one can strictly improve both the \textit{SOP} and the \textit{ESR} compared to a non-precoded Gaussian input.

%%%%%%%%%%%%%%%%%%%%%%%%%%%%%%%%%%%%%%%%%%%%%%%%%%%
%%                                               %%
%%                  CONCLUSION                   %%
%%                                               %%
%%%%%%%%%%%%%%%%%%%%%%%%%%%%%%%%%%%%%%%%%%%%%%%%%%%
\section{Conclusion}
\label{sec:conclusion}
In this paper, we study the optimization of three secrecy metrics for wireless channels under the assumption that only channel statistics are available at the transmitter. We show that, when the legitimate channel is \textit{uniformly less noisy} than the eavesdropper channel, both the Secrecy Outage Probability (\textit{SOP}) and the Ergodic Positive Secrecy Rate (\textit{EPSR}) are maximized by a non-precoded Gaussian input. The same input distribution also maximizes the Ergodic Secrecy Rate (\textit{ESR}) under the weaker condition that the main channel is \textit{less noisy on average} than the eavesdropper channel.

Reducing the problem to SIMOME channels still yields the same optimal input distribution for the \textit{SOP} and the \textit{EPSR}, regardless of the channel ordering. However, in the absence of assumptions on the number of transmit antennas or on channel ordering, no unique general input distribution maximizes the \textit{ESR}. This implies that \textit{ESR} optimization must be carried out on a case-by-case basis, depending on the specific statistical characteristics of the channels.

In MIMOME channels, assumptions on channel ordering render certain functions involved in the optimization of the secrecy metrics concave, thereby enabling tractable optimization. In SIMOME channels, the optimization of the \textit{SOP} and the \textit{EPSR} relies on the fact that a non-precoded Gaussian input achieves the maximum instantaneous secrecy rate whenever Bob's channel is stronger than Eve's, and on the fact that SIMO channels can be totally ordered. Scenarios in which Eve's channel is stronger are irrelevant in this context, since both the \textit{SOP} and the \textit{EPSR} only capture how large the secrecy rate is when it is positive, rather than how negative it becomes otherwise. For some channels, the non-precoded Gaussian input is not optimal, which shows that the channel ordering assumptions cannot be simply relaxed. Characterizing more general conditions under which the non-precoded Gaussian input is optimal remains an open problem.

%%%%%%%%%%%%%%%%%%%%%%%%%%%%%%%%%%%%%%%%%%%%%%%%%%%
%%                                               %%
%%                 BIBLIOGRAPHIE                 %%
%%                                               %%
%%%%%%%%%%%%%%%%%%%%%%%%%%%%%%%%%%%%%%%%%%%%%%%%%%%

\end{document}